\documentclass[aps,prd,preprintnumbers,superscriptaddress,twocolumn,nofootinbib]{revtex4}%,showpacs %nofootinbib %preprintnumbers,
\usepackage[dvips]{graphicx}
\usepackage{bm,latexsym,amsmath,amssymb,amsfonts,mathrsfs}
\usepackage{hyperref}
\usepackage[export]{adjustbox}
\usepackage{color}
\usepackage[normalem]{ulem}
\input{colordvi.tex}
\newcommand{\bea}{\begin{eqnarray}}
\newcommand{\eea}{\end{eqnarray}}
\newcommand{\nn}{\nonumber\\}

\hypersetup{colorlinks=true,
	breaklinks=true,
	pdfstartview=Fit,
	linkcolor=blue,
	citecolor=blue,
	urlcolor=blue}

\begin{document}

\preprint{KOBE-COSMO-22-11}
\title{Superheavy Dark Matter Production from Symmetry Restoration First-Order Phase Transition During Inflation}

\author{Haipeng An}
\email{anhp@mail.tsinghua.edu.cn}
\affiliation{Department of Physics, Tsinghua University, Beijing 100084, China}
\affiliation{Center for High Energy Physics, Tsinghua University, Beijing 100084, China}

\author{Xi Tong}
\email{xtongac@connect.ust.hk}
\affiliation{Department of Physics and Jockey Club Institute for Advanced Study,
	Hong Kong University of Science and Technology, Hong Kong}

\author{Siyi Zhou}
\email{siyi@people.kobe-u.ac.jp}
\affiliation{Department of Physics, Kobe University, Kobe 657-8501, Japan}

\begin{abstract}
We propose a scenario where superheavy dark matter (DM) can be produced via symmetry restoration first-order phase transition during inflation triggered by the evolution of the inflaton field. The phase transition happens in a spectator sector coupled to the inflaton field. During the phase transition, the spectator field tunnels from a symmetry-broken vacuum to a symmetry-restored vacuum. The massive particles produced after bubble collisions are protected against decaying by the restored symmetry and may serve as a DM candidate in the later evolution of the Universe. We show that the latent heat released during the phase transition can be sufficient to produce the DM relic abundance observed today. In addition, accompanied with the super heavy DM, this first-order phase transition also produces gravitational waves detectable via future gravitational wave detectors.
\end{abstract}

\maketitle
%%%%%%%%%%%%%%%%%%%%%%%%%%%%%%%%%%%%%%%%%%%%%%%%%%%%%%%%%%%%%%%%%%%%%%

\section{Introduction}
Despite the fact that most of the matter in our universe is dark, the particle nature of DM and its production mechanism remains unknown. The most popular DM production scenario is the so-called Weakly Interacting Massive Particles (WIMPs) with the freeze-out mechanism. This type of DM starts out in thermal equilibrium in the early universe. As the universe expands, the interaction rate of WIMPs drops below the expansion rate of the Universe, and the WIMPs can no longer maintain thermal equilibrium with ordinary matter. They then exit the thermal bath and become thermal relic today. Since the annihilation cross section is inversely proportional to the mass square of the DM particles, DM particles which are too heavy will drop out of the thermal equilibrium too early, resulting in the overproduction of DM. This sets an upper limit on the mass of the DM particle, which is around $10^{5}$ GeV~\cite{Griest:1989wd}. 

DM heavier than $10^{5}$~GeV can be produced non-thermally via the freeze-in mechanism~\cite{McDonald:2001vt,Hall:2009bx,Bernal:2017kxu,Goudelis:2018xqi,Chowdhury:2018tzw,Bhattacharyya:2018evo,Bernal:2018qlk,Chung:1998rq,Chung:1998ua,Giudice:2000ex}. Alternative scenarios are also proposed, such as Planckian Interacting DM~\cite{Garny:2015sjg,Garny:2017kha,Hashiba:2018iff,Haro:2018zdb,Hashiba:2018tbu}, SUPERWIMP~\cite{Feng:2010gw}, FIMP~\cite{Hall:2009bx} etc. See \cite{Ema:2019yrd,Ahmed:2020fhc,Gross:2020zam,Ahmed:2021fvt,Xue:2021jyj,Ema:2021fdz,Clery:2022wib,Wang:2022ojc} for relevant recent developments along this line. For a review of the non-thermal DM production mechanisms, see Ref.~\cite{Carney:2022gse}.
Superheavy DM can also be produced during either inflation \cite{Parker:1969au,Ford:1986sy,Li:2019ves,Li:2020xwr,Ling:2021zlj} or the transition between inflation and the subsequent evolution of the universe. The particle production mechanism is mostly related to the non-trivial evolution of the cosmological background. Particles can be produced from vacuum through time-dependent Bogoliubov transformations computable via the Stokes-line method \cite{Sou:2021juh}. See \cite{Li:2019ves,Hashiba:2020rsi,Hashiba:2021npn,Yamada:2021kqw,Basso:2021whd} for recent works and \cite{Ford:2021syk} for a recent review on this topic.
%

%In contrast to the freeze-out mechanism, superheavy DM is usually non-thermally prepared via the so-called the freeze-in mechanism~\cite{McDonald:2001vt,Hall:2009bx,Bernal:2017kxu,Goudelis:2018xqi,Chowdhury:2018tzw,Bhattacharyya:2018evo,Bernal:2018qlk,Chung:1998rq,Chung:1998ua,Giudice:2000ex}~\footnote{It is also possible to produce thermal superheavy DM within the freeze out mechanism~\cite{Berlin:2017ife,Kramer:2020sbb}. }.Recently, 

Superheavy DM can be produced during inflation via the quantum fluctuation. However, the amount of DM particles may not be sufficient to explain the relic abundance of the DM today. This is because the inflationary spacetime resembles that of a de Sitter (dS) space, and the dS-invariant production number density of heavy DM particles is suppressed exponentially by the factor $e^{-2\pi\sqrt{\frac{m_{\rm DM}^2}{H^2}-\frac{9}{4}}}$ once $m_{\rm DM}\gg H$, where $H$ is the Hubble expansion rate\footnote{Unless specified, $H$ in this paper always stands for the Hubble parameter during inflation.}. 
%\sim e^{-m/T_{dS}}$, where $T_{dS}=\frac{H}{2\pi}$ is the dS background temperature.

%If such kind of superheavy DM is solely produced from vacuum during inflation, the amount of DM particles may not be sufficiently enough to explain the relic abundance of the DM today. This is because the inflationary spacetime resembles that of a de Sitter (dS) space, and the dS-invariant production number density of the massive DM particles is severely suppressed by the Boltzmann factor $e^{-2\pi\sqrt{\frac{m^2}{H^2}-\frac{9}{4}}}\sim e^{-m/T_{dS}}$, where $T_{dS}=\frac{H}{2\pi}$ is the dS background temperature. 

In this work, we propose a much more efficient DM production mechanism, by introducing a symmetry restoration first-order phase transition (SRFOPT) during inflation. Motivated by the fact that the inflaton $\phi$ traverses a large distance in field space~\cite{Lyth:1996im}, the inflaton may encounter non-trivial features on the scalar manifold. We consider a massive spectator scalar field $\sigma$ weakly coupled to the rolling inflaton. The inflaton rolling off the potential $V(\phi,\sigma)$ triggers a SRFOPT in the $\sigma$ direction. When the  phase transition happens, the energy difference between the true vacuum and false vacuum is injected into expanding bubble walls. The subsequent collisions of bubble walls dissipate the energy into $\sigma$ particles, producing observable gravitational wave (GW) signals simultaneously. These $\sigma$ particles end up in a symmetric vacuum and are thus protected from decaying, making them perfect superheavy DM candidates. 

Tthe signatures of the accompanying GW signal are also important as they provide further information about the phase transition. We point out that this is only when the symmetry restoration phase transition is \textit{first-order}. Second order phase transitions (such as the gravitational misalignment mechanism \cite{Babichev:2020xeg,Laulumaa:2020pqi,Karam:2020rpa,Borah:2020ljr,Ramazanov:2021eya}) can generate gravitational wave by domain wall production. 
The GW signatures from phase transition during inflation have been studied in \cite{Jiang:2015qor,Wang:2018caj,Li:2020cjj,An:2020fff,An:2022cce} (see also \cite{Sugimura:2011tk} for the GW signature from phase transition before inflation), and their unique oscillation signatures can help us distinguish them from those produced during post inflationary evolutions. As a result, this GW signature can serve as an indirect probe of superheavy DM produced during inflation.

\begin{figure}[t] 
	\centering 
	\includegraphics[width=8cm]{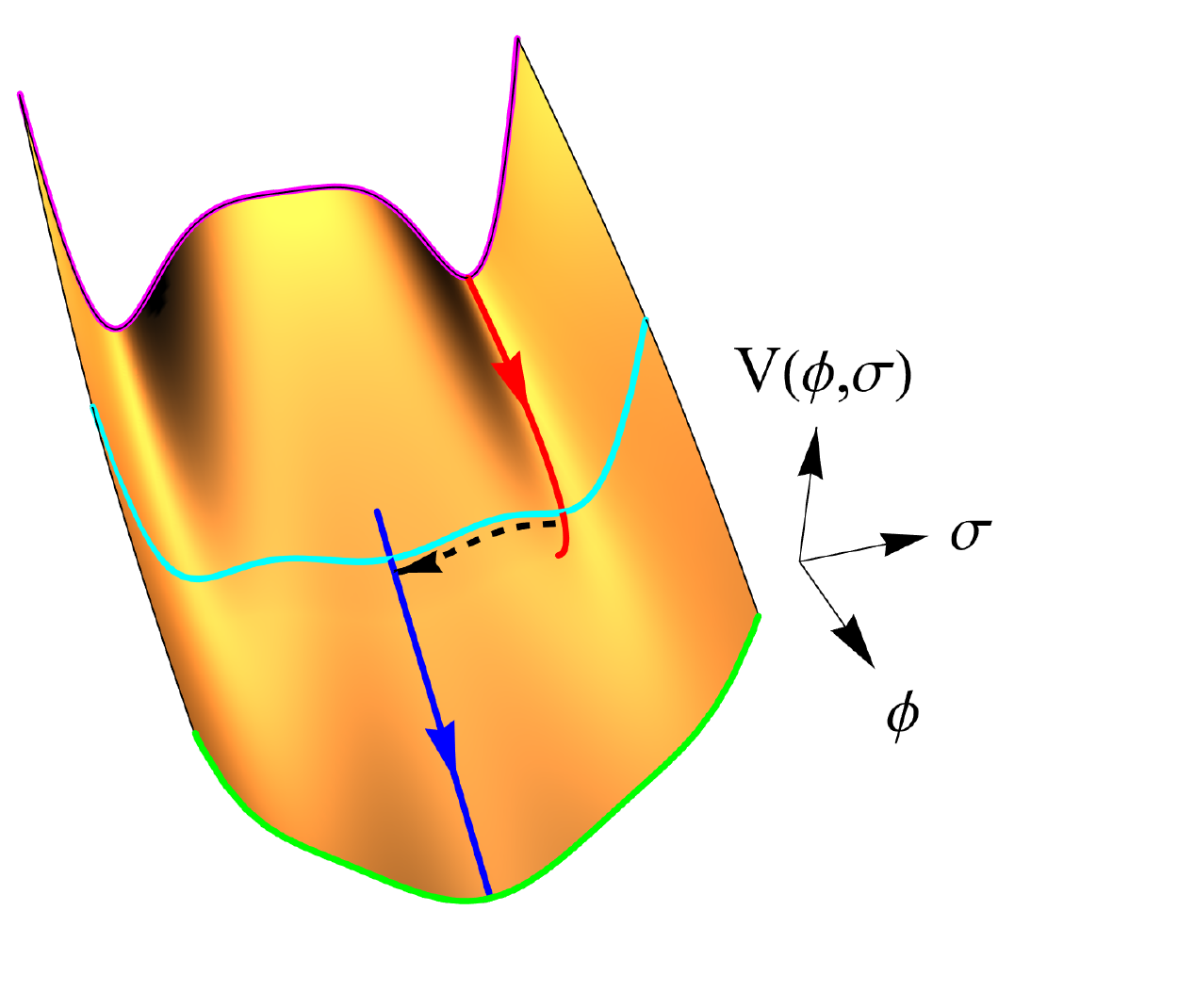}  
	\caption{The potential shape $V(\phi,\sigma)$ and the field trajectory. The magenta, cyan and green curves show the plane sections along the $\sigma$ direction. The inflaton rolls down along the $\phi$ direction due to a slow-roll potential. Initially, the field follows the false vacuum (red). Then as the true vacuum (blue) forms and becomes lower than the false vacuum, the $\sigma$ field tunnels through the barrier (black dashed line) and triggers the SRFOPT. 
	} \label{potential3D}
\end{figure}

In the rest of the paper, we illustrate this idea with a simple but viable model. The spectator sector we consider is composed by a real scalar field $\sigma$, and the effective action is written as
\begin{equation}
	S \equiv \int d^4x \sqrt{-g} \bigg[-\frac{1}{2}(\partial\phi)^2-\frac{1}{2}(\partial\sigma)^2-U(\phi,\sigma)\bigg]~,
\end{equation}
where $\phi$ is the inflaton field. The potential takes the form 
\begin{equation}
	U(\phi, \sigma)\equiv V(\phi, \sigma)+V_{\rm sr}(\phi) ~,\label{Upotential}
\end{equation}
where
\begin{equation}
	V(\phi, \sigma)\equiv\frac{1}{2} \mu_{\rm eff}^2(\phi) \sigma^{2}+\frac{\lambda}{4} \sigma^{4}+\frac{1}{8 \Lambda^{2}} \sigma^{6}~,\label{MainModel}
\end{equation}
and $V_{\rm sr}(\phi)$ is the usual slow-roll potential of the inflaton sector in the slow-roll inflation. Notice the $Z_2$ symmetry $\sigma\leftrightarrow -\sigma$ of the Lagrangian. We assume the $\sigma$ sector has an energy density subdominant to that of the inflaton, $i.e.$, $V(\phi,\sigma)\ll V_{\rm sr}(\phi)$. As mentioned before, the rolling $\phi(t)$ field background introduces a time dependent effective mass of the $\sigma$ field,
\begin{equation}
	\mu_{\rm eff}^2(\phi)\equiv \mu^{2} - c^{2} \phi^{2}~,
\end{equation}
triggering the symmetry restoration phase transition in the $\sigma$ sector (see FIG.~\ref{potential3D} for a cartoon illustration, assuming $\phi$ rolling down to $0$ from a large value during inflation).

In Sec.~\ref{ModelSection}, we describe the details of the phase transition this model. Then in Sec.~\ref{DMSection}, we analyze the evolution of the $\sigma$ particle number density and calculate the DM relic abundance. We move on to the accompanying gravitational wave signals in Sec.~\ref{GWSection}. We sumarize and give outlooks in Sec.~\ref{ConclusionSection}.

\section{First-order phase transition induced by the evolution of the inflaton}\label{ModelSection}

%Considering the following model with two fields,
%\begin{equation}
%	S \equiv \int d^4x \sqrt{-g} \bigg[-\frac{1}{2}(\partial\phi)^2-\frac{1}{2}(\partial\sigma)^2-U(\phi,\sigma)\bigg]~.
%\end{equation}
%where $\phi$ is the inflaton and $\sigma$ is a massive spectator field. The potential takes the form
%\begin{equation}
%	U(\phi, \sigma)\equiv V(\phi, \sigma)+V_{\rm sr}(\phi) ~,\label{Upotential}
%\end{equation}
%where
%\begin{equation}
%	V(\phi, \sigma)\equiv\frac{1}{2} \mu_{\rm eff}^2(\phi) \sigma^{2}+\frac{\lambda}{4} \sigma^{4}+\frac{1}{8 \Lambda^{2}} \sigma^{6}~,\label{MainModel}
%\end{equation}
%and $V_{\rm sr}(\phi)$ is the usual slow-roll potential of the inflaton sector in the slow-roll inflation. We assume the $\sigma$ sector has an energy density subdominant to that of the inflaton, $i.e.$, $V(\phi,\sigma)\ll V_{\rm sr}(\phi)$. As mentioned before, the rolling $\phi(t)$ field background introduces a time dependent effective mass of the $\sigma$ field,
%\begin{equation}
%	\mu_{\rm eff}^2(\phi)\equiv c^{2} \phi^{2}-\mu^{2}~,
%\end{equation}
%triggering the inverse phase transition in the $\sigma$ sector (see FIG.~\ref{potential3D} for a cartoon illustration).

%We will require the inverse phase transition finishes within a time scale much shorter than the Hubble scale. 
%The condition whether first-order phase transition can be completed during inflation has been estimated in \cite{An:2020fff}. 
%And the condition whether inverse first-order phase transition can happen is the same. 
Denoting the typical mass scale of the $\sigma$ particles as $m_\sigma\sim\mu$, the bubble nucleation rate per physical volume of the universe can be written as 
\begin{equation}
	\frac{\Gamma}{V_{\rm phys}} = \mathcal O(1) \times  m_{\sigma}^4 e^{-S_4}~,
\end{equation}
where $S_4$ is the classical action of the bounce solution. We define the parameter $\beta$ to characterize how fast the SRFOPT happens
\begin{equation}
	\beta\equiv -\frac{d S_4(t)}{dt}= \mathcal{O}(100)\times\left|\frac{\dot{\phi}}{\phi-\frac{\mu^2}{c^2\phi}}\right| ~.\label{betaDef}
\end{equation}
In order that the phase transition can finish in time, we require the $\beta$ parameter to be sufficiently large. The typical value of $\beta$ we consider is of order $\beta \sim 5 H$. The universe should be filled with true vacuum bubbles before significant inflationary expansion tears them apart. This condition indicates that \cite{An:2020fff}
\begin{equation}\label{eq:condition1}
	m_\sigma^4\gg \beta^4~.
\end{equation} 
On the other hand, the requirement that the energy density of the spectator sector is subdominant yields
\begin{equation}
	m_\sigma^4\sim V(\phi,\sigma)\ll V_{\rm sr}(\phi)\sim M_{\rm pl}^2 H^2 ~.
\end{equation}

During the SRFOPT, the bubbles collide into each other, producing a large amount of $\sigma$ particles that thermalize quickly due to mutual interactions. These massive $\sigma$ particles are formed in the $Z_2$-symmetric phase. Therefore, the $\sigma$ particles are stable and can be the DM candidate. 
%Considering that its only non-gravitational coupling is with the inflaton field, which has been depleted after reheating, these $\sigma$ particles can play the role of DM we observe today.

%In the following, we take $m_\sigma \sim \mathcal{O}(10^2$-$10^3) H$ as our benchmark parameter choice and try to see if the $\sigma$ field can account for the relic abundance of DM of our universe. 
The energy of the $\sigma$ particles comes from the colliding bubble walls, which originates from the latent heat
\begin{equation}
	L\equiv \gamma_{\rm PT} m_\sigma^4
\end{equation}
of the SRFOPT. Here $\gamma_{\rm PT}$ is a dimensionless factor that can, in principle, be computed from the model (\ref{MainModel}) itself. However, it is more convenient to consider it as an input parameter for generic potentials that take forms different from (\ref{MainModel}). Since the tunneling processes mainly happen at critical points where different terms in the potential $V(\phi,\sigma)$ are of comparable sizes, $\gamma_{\rm PT}$ will not be far from unity in general. 

One way to estimate the upper bound of $\gamma_{\rm PT}$ is to look at the potential $V(\phi,\sigma)$ and get the maximum value of the difference between the true vacuum and the false vacuum. In the symmetry-broken phase, the potential of the $\sigma$ field has three minima, while at the symmetric phase, the potential of the $\sigma$ field has only one minimum at $\sigma=0$. The critical case is when the two minima away from $\sigma=0$ disappear, which leads to the condition $\lambda^2=3\mu_{\rm eff}^2/\Lambda^2$. The minima are located at $\sigma_{\rm false}=\pm\sqrt{-2\lambda/3}\Lambda$ and $\sigma_{\rm true} = 0$. The difference of the vacuum energy is thus $-\lambda^3\Lambda^4/27$. So we have 
\begin{align}
	\gamma_{\rm PT} m_{\rm \sigma}^4 \lesssim -\lambda^3\Lambda^4/27~.
\end{align}
For instance, a parameter choice $\lambda=-1$, $\Lambda=2 m_\sigma$ leads to $\gamma_{\rm PT}\lesssim 0.6$.

%Therefore, after thermalization, the average energy of the $\sigma$ particles is comparable to $m_\sigma$. As the universe expands, the kinetic energy of the $\sigma$ particles will decrease, making them more and more non-relativistic. 
%Hence, we will adopt the \textit{non-relativistic} approximation in the  calculation of the evolution of the $\sigma$ particles. 

%
%Hence for simplicity, we will adopt the \textit{non-relativistic} approximation in the following calculations. The reason why we take this approximation is threefold. First, the estimated upper bound $\gamma_0\lesssim 0.6$ corresponds to only mildly relativistic $\sigma$ particles produced immediately after bubble collisions. Second, the bubble-to-particle conversion may not be completely efficient and there will be a reduction factor for the energy density of the thermal bath. Third, as the universe expands, the kinetic energy of the $\sigma$ particles will decrease, making them more and more non-relativistic. In fact, we will see in Sect.~\ref{DMSection} that the non-trivial evolution of comoving particle number density due to $\sigma\sigma\sigma\sigma\to\sigma\sigma$ scattering and the freeze-out process do not happen until the thermal bath becomes non-relativistic.

Since the during the phase transition the $\sigma$ particles are tightly coupled to each other, and $m_\sigma\gg \beta$, it is reasonable to assuming that the marginally relativistic $\sigma$ particles immediately thermalize after the phase transition. The distribution function of the $\sigma$ particles is then given by the Maxwell-Boltzmann distribution
\begin{equation}\label{MBdistribution}
	f(E,T(t)) = e^{-E/T(t)},~ E(p)=\sqrt{m_\sigma^2+\mathbf p^2}~. 
\end{equation}
The initial temperature $T_{\rm PT}\equiv T(t_{\rm PT})$ can then be determined from the latent heat by 
\begin{align}\label{latentheat}
	L\simeq\gamma_{\rm PT} m_\sigma^4 =\int \frac{d^3\mathbf p}{(2\pi)^3} E(p) f(E,T_{\rm PT})~.
\end{align}
The initial number density of $\sigma$ particles is
\begin{equation}
	n_\sigma(t_{\rm PT})=\int \frac{d^3\mathbf p}{(2\pi)^3} f(E,T_{\rm PT})~.\label{initialNumberDensity}
\end{equation}
Later, as the universe expands, the temperature $T(t)$ drops and the energy density of the thermal bath decays as
\begin{equation}
	\rho(t)=\gamma(t) m_\sigma^4 =\int \frac{d^3\mathbf p}{(2\pi)^3} E(p) f(E,T(t))~.
\end{equation}
The $\sigma$ particles become approximately non-relativistic when $\langle E\rangle\simeq 2m_\sigma$ at $t_*$ with
\begin{equation}
	\gamma_*\equiv\gamma(t_*)\simeq 0.01~.
\end{equation}
After $t_*$, there will be a slight imbalance in the reaction rates and the distribution starts to deviate from the equilibrium result (\ref{MBdistribution}). For now, let us first neglect such a process. Later we will see that it does not change the final relic abundance significantly. 

According to the Planck 2018 \cite{Planck:2018nkj} and BAO \cite{Planck:2018lbu} result, the relic abundance of the DM is
\begin{equation}
	\Omega_{\sigma} h^2 =  0.11923 ~.
\end{equation}
The absolute value of the energy of DM today is 
\begin{align}\nonumber
	\rho^{(0)}_{\rm DM} = \Omega_{\sigma} \rho_0 &= \frac{0.11923}{0.68^2} ( 3 M_{\rm pl}^2{H}_0^2)\hbar ^2 {c}^4 \\
	&= (1.76\times 10^{-12}{\rm GeV})^4~.\label{PlanckBAODMEnergyDensity}
\end{align}
where $\rho_0$ is the critical energy density of the universe today. 
If the SRFOPT happened $N_*$ e-folds before the end of inflation, the energy density of DM today can be written as
%During the SRFOPT, if the phase transition happens at $N_0 = 18.1$ e-folds before the end of inflation, the energy density of the $\sigma$ particles today will be close to that of the observational value (\ref{PlanckBAODMEnergyDensity}):
\begin{align}\label{eq:rhosigma0}
	\rho_\sigma^{(0)} &\sim \gamma_* \times m_\sigma^4 e^{-{3(N_{\rm today}+N_{\rm PT}-\frac{1}{4}\ln(\gamma_{\rm PT}/\gamma_*))}}\ ,
%	&\simeq (1.76\times 10^{-12}{\rm GeV})^4\simeq \rho_{\rm DM}  ~.\label{DMRelicAbundanceEstimation}
\end{align}
%\xt{Check numerics}
where $N_{\rm today}$ is the e-folds the universe expands from the end of inflation to today. Then, by requiring $\rho_\sigma^{(0)} = \rho_{\rm DM}^{(0)}$, we can calculate $N_{\rm PT}$ for a given scenario of the universe history. For instance, if we assume $H = 10^{12}~{\rm GeV}$ and require that the reheating process finished within one e-fold, we can get that $N_{\rm today} \approx 65$ and thus obtain $N_{\rm PT} \approx 18$. 

%where we have assumed typical values $\gamma_* \simeq 0.01$, $\gamma_0\simeq 0.6$, $m_\sigma \simeq 2000H$, $N_0 \simeq 18.1$ and $H\simeq 10^{12}$ GeV. The e-folding number from the end of inflation to today is taken to be around $N_{\rm today}=64.5$ assuming instantaneous reheating. As a result, the $\sigma$ particles produced in an SRFOPT during inflation can indeed account for the DM today. 

\section{DM number density evolution after SRFOPT}\label{DMSection}

\begin{figure}[t] 
	\centering 
	\includegraphics[width=7cm]{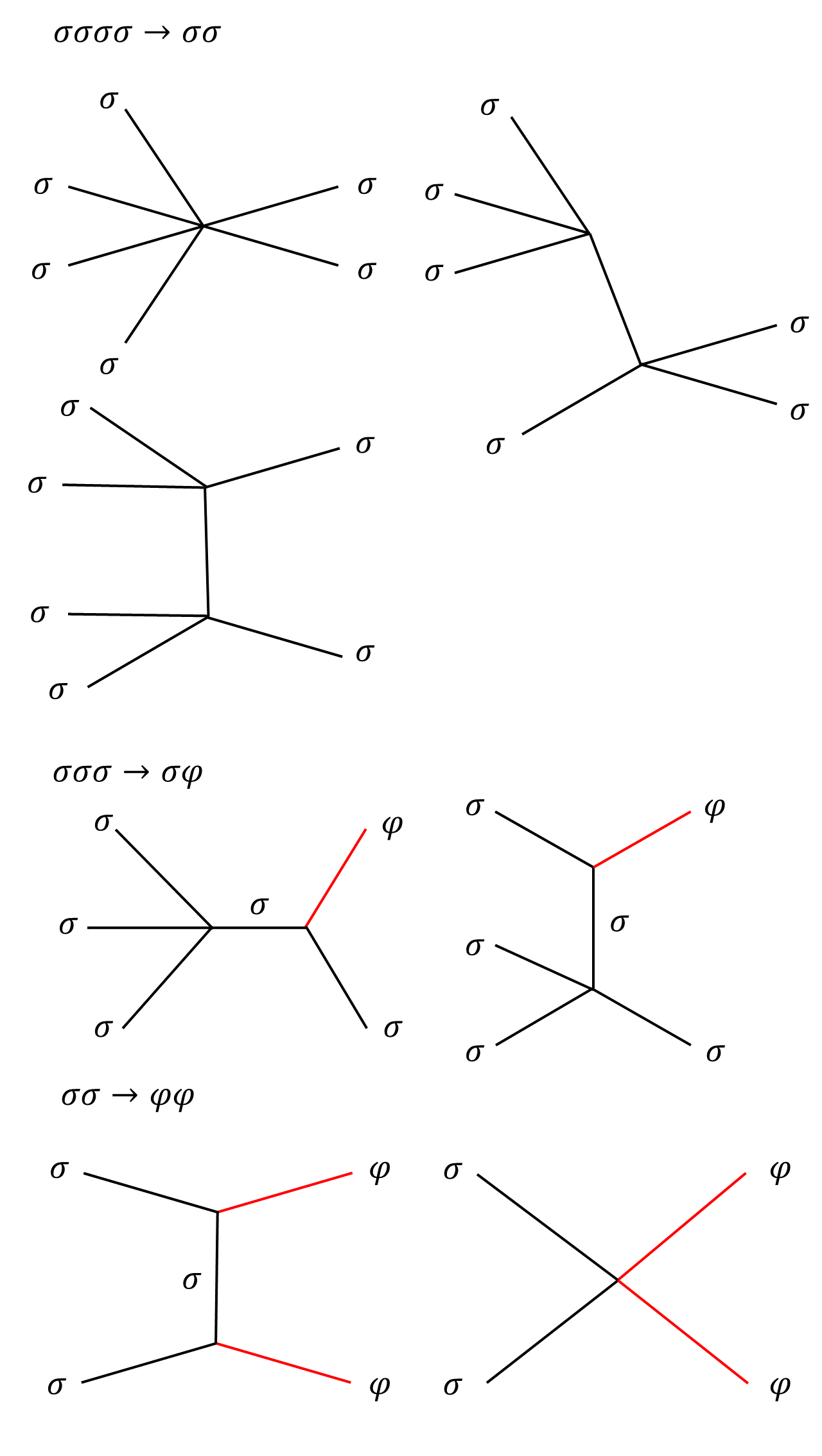}  
	\caption{The diagrams that contribute to the decreasing in the number of the $\sigma$ field.
	} \label{2to2}
\end{figure}

At the end of the SRFOPT, the $\sigma$ particles thermalize due to efficient $\sigma\sigma\to\sigma\sigma$ scattering. This can be seen from estimating the interaction rate
\begin{equation}
	\Gamma=n_\sigma(t_{\rm PT}) \sigma_{2\to 2}v\sim \gamma_{\rm PT} m_\sigma^3 \cdot\frac{\lambda^2}{m_\sigma^2}\gg H~.
\end{equation}
In a marginally relativistic thermal equilibrium, the $\sigma$ self-interactions lead to balanced $\sigma\sigma\sigma\sigma\to\sigma\sigma$ and $\sigma\sigma\to\sigma\sigma \sigma\sigma$ scatterings. Yet as the universe expands, the average energy of the $\sigma$ particles drops below the $\sigma\sigma\rightarrow\sigma\sigma\sigma\sigma$ threshold, leading to the imbalance between these two processes. In addition, 
the number density of $\sigma$ particles can also reduce through the process of producing the inflaton particles. 
More explicitly, the inflaton field can be decomposed into the homogeneous background $\bar\phi$ and the perturbation part $\varphi$,
\begin{align}
	\phi(x) = \bar\phi(t) + \varphi (t,\mathbf x)~.
\end{align}
The potential term can thus be expanded as
\begin{align}
	\nonumber V(\phi, \sigma)  =  &\frac{1}{2}\mu_{\rm eff}^{2}(\bar{\phi})\sigma^{2}\\
	&+c^2\bar\phi\varphi\sigma^{2} + \frac{1}{2}c^2\varphi^{2}\sigma^{2}  +\frac{\lambda}{4} \sigma^{4}+\frac{1}{8 \Lambda^{2}} \sigma^{6} ~.  \label{eq:Vphisigma}
\end{align}
After estimating the interaction rates, we find that the Feynman diagrams in Fig.~\ref{2to2} give the main contributions to reduce the comoving DM number density. Thus the decrease of the number density of the $\sigma$ particle can be described using the following Boltzmann equation,
\begin{align}
	a^{-3} \frac{d(n_\sigma a^3)}{dt}=C_{\sigma\sigma\sigma\sigma\rightarrow\sigma\sigma}+C_{\sigma\sigma\rightarrow\varphi\varphi}+C_{\sigma\sigma\sigma\rightarrow\sigma\varphi}~.\label{thermalization}
\end{align}
%The collision terms come from the dominant channels shown in FIG.~\ref{2to2}. 
The detailed computations for the collision terms are given in Appendix~\ref{DMReductionRateAppendix}. At $t=t_{\rm PT}$, the $\sigma$ particle number density $n_\sigma(t_{\rm PT})$ is given by Eqs.~(\ref{latentheat}) and (\ref{initialNumberDensity}). Right after the phase transition, the evolution of $n_\sigma(t)$ follows that in a thermal equilibrium, where the comoving number density is kept as a constant. The deviation from equilibrium starts at $t=t_*$, with $\gamma_*=0.01$. Therefore, the subsequent evolution of $n_\sigma(t)$ can be solved from (\ref{thermalization}) with an initial condition $n_\sigma(t_*)=\gamma_* m_\sigma^3/2$. As shown by Eq.~(\ref{Boltzmann6sigma}), in the non-relativistic regime, $C_{\sigma\sigma\sigma\sigma\rightarrow\sigma\sigma}$ can be estimated as
\begin{equation}
C_{\sigma\sigma\sigma\sigma\rightarrow\sigma\sigma} \sim -  \frac{n_\sigma^4}{m_\sigma^8} \ .
\end{equation}

To estimate $C_{\sigma\sigma\sigma\rightarrow \sigma\varphi}$, the three-point coupling vertex of $\sigma^2 \varphi$ can be read from Eq.~(\ref{eq:Vphisigma}) as $c^2 \bar\phi$. During phase transition, the $c^2 \bar\phi^2$ is at the order of magnitude of $m_\sigma^2$. Therefore, the 3pt coupling can be estimated as $m_\sigma^2/\bar\phi \sim m_\sigma^2/M_{\rm pl}$. Therefore, we have
\bea
C_{\sigma\sigma\sigma\rightarrow\sigma\varphi} &\sim& -\frac{ n_\sigma^3}{(2 m_\sigma)^3 m_\sigma^4} \left( \frac{m_\sigma^2}{M_{\rm pl}} \right)^2 \ ,
\eea
where the approximation $L \approx n_\sigma m_\sigma$ is used. 

For the annihilation process, $C_{\sigma\sigma\rightarrow\varphi\varphi}$ can be estimated as 
\bea
C_{\sigma\sigma\rightarrow\varphi\varphi} \sim \frac{c^4 n_\sigma^2}{(2 m_\sigma)^2} \ .
\eea
During SRFOPT, we have $c^2 \sim m_\sigma^2/M_{\rm pl}^2$. Therefore, right after the completion of the SRFOPT we have 
\bea
C_{\sigma\sigma\rightarrow\varphi\varphi} \sim \left(\frac{L}{\rho_{\rm inf}}\right)^2 H^4 \ .
\eea
Now, we can calculate the evolution of the comoving number density. We define $Y_\sigma(t) = n_\sigma(t) a^3(t)/ n_\sigma(t_*) a^3(t_*)$, neglecting all the order one factors, the evolution equation of $Y_\sigma$ can be written as
\bea\label{eq:Y}
\frac{dY_\sigma}{dt} &\sim& - \frac{Y_\sigma^4 n_\sigma^3(t_*)}{m_\sigma^8} e^{ - 9 H (t - t_*)} - \frac{Y_\sigma^3 n_\sigma^2(t_*)}{m_\sigma^3 M_{\rm pl}^2} e^{-6H(t-t_*)} \nn
&& - \frac{Y_\sigma^2 n_\sigma(t_*) m_\sigma^2 }{M_{\rm pl}^4} e^{- 3 H(t-t_*)} \ ,
\eea
with the initial condition $Y_\sigma(t_*) = 1$. The integration of $t$ will produce a factor of $1/H$ for each term on the righthand side of Eq.~(\ref{eq:Y}). For the second term, this contribution is equivalent to decrease $Y_\sigma$ by a factor of about
\bea
\frac{n_\sigma^2(t_*)}{m_\sigma^3 M_{\rm pl}^2 H} \sim \frac{\gamma_*^2 m_\sigma^3}{M_{\rm pl}^2 H} \sim \frac{\gamma_*^2 H}{m_\sigma} \frac{L}{\rho_{\rm inf}} \ .
\eea  
From the condition (\ref{eq:condition1}), we require $m_\sigma \gg \beta \gg H$ for the SRFOPT to complete. Furthermore, the non-relativistic condition gives $\gamma_* \approx 0.01$. We also require $L \ll \rho_{\rm inf}$ for the spectator sector to be subdominant. Therefore, the second term on the righthand side of Eq.~(\ref{eq:Y}) can only produce a negligible effect on the evolution of $Y_\sigma$. Similar argument shows that the effect from the third term on the righthand side of Eq.~(\ref{eq:Y}) is also negligible. Therefore, (\ref{eq:Y}) can be simplified as 
\bea\label{eq:dT}
\frac{dY_\sigma}{dt} &=& - \frac{ {\cal C} Y_\sigma^4 n_\sigma^3(t_*)}{m_\sigma^8} e^{ - 9 H (t - t_*)} \ ,
\eea
where the ${\cal C}$ is a numerical factor collecting the information of the couplings, symmetry factors, and phase space factors in the calculation of the collision term. Depending on the detailed choice of the model parameters, ${\cal C}$ varies from ${\cal O}(1)$ to ${\cal O}(0.001)$. The details are presented in Appendix~\ref{DMReductionRateAppendix}. 
Thus, we have
\bea
Y_\sigma(\infty) = \left( 1 + \frac{  {\cal C} n_\sigma^3(t_*) }{3 m_\sigma^8 H} \right)^{-1/3} = \left(1 + \frac{ {\cal C}\gamma_*^3 m_\sigma}{24 H}\right)^{-1/3}  . 
\eea 
Therefore the formula for today's DM energy density (\ref{eq:rhosigma0}) should be written as 
\bea
\rho_\sigma^{(0)} = \gamma_*  m_\sigma^4 Y_\sigma(\infty) e^{-{3(N_{\rm today}+N_{*}-\frac{1}{4}\ln(\gamma_{\rm PT}/\gamma_*))}}\ .
\eea

\begin{figure}[t] 
	\centering 
	\vspace*{0.3cm}
	\includegraphics[width=8.5cm]{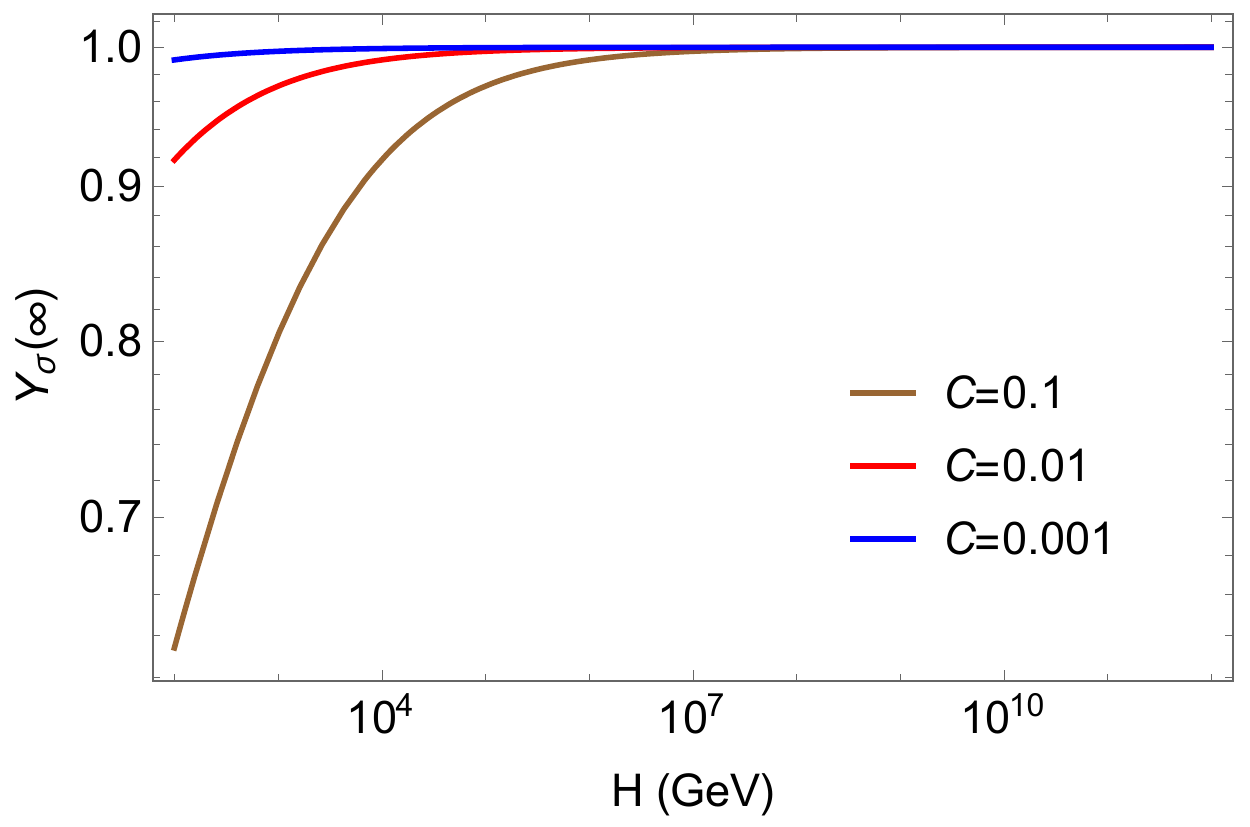}  
	\caption{$Y_\sigma(\infty)$ as a function of $H$ for different values of ${\cal C}$. Here the value of $L/\rho_{\rm inf}$ is fixed to be $0.1$. } \label{fig:Ysigma}
\end{figure}

In next section, we will discuss the production of the GWs during SRFOPT. The strength of the GWs today is proportional to $(L/\rho_{\rm inf})^2$. Thus, we rewrite $Y_\sigma(\infty)$ as a function of $L/\rho_{\rm inf}$,
\bea\label{eq:Yinfty}
Y_\sigma(\infty) \approx \left( 1 + \frac{ \gamma_*^3 {\cal C}}{24} \left(\frac{M_{\rm pl}}{ H}\right)^{1/2} \left(\frac{L}{\rho_{\rm inf}}\right)^{1/4} \left(\frac{3}{\gamma_{\rm PT}}\right)^{1/4}\right)^{-1/3} .
\eea
With $L/\rho_{\rm inf}$ fixed to 0.1, $\gamma_{\rm PT}$ fixed to 1, $Y_\sigma(\infty)$ as functions of $H$ for different values of ${\cal C}$ are shown in Fig.~\ref{fig:Ysigma}. One can see that the $\sigma\sigma\sigma\sigma\rightarrow\sigma\sigma$ process does not significantly reduce the comoving number density of $\sigma$ particle as long as $H > 100$ GeV. 

\section{Gravitational Wave Signals}\label{GWSection}

GWs often serve as indirect probes to superheavy DM. Superheavy DM production from first-order phase transition after inflation and its complementary GW signatures has already been studied extensively in the literature \cite{Baker:2019ndr,Chway:2019kft,Marfatia:2020bcs,Baldes:2020kam,Azatov:2021ifm,Bian:2021vmi,Freese:2022qrl,Caldwell:2022qsj}. Here we discuss the GW properties related to superheavy DM produced via first-order phase transition during inflation. The strength of the GW signal strongly spends on the reheating scenario after inflation. 

\subsection{Instantaneous reheating scenario}

In scenarios such as parametric resonant preheating, the energy stored in the inflaton potential thermalized within e-fold~\cite{Traschen:1990sw,Kofman:1997yn,Parry:1998pn,Sornborger:1998mh}. In this case, the frequency of the GWs observed today can be calculated by counting the redshifts at various eras of the evolution of the Universe that~\cite{An:2020fff}
\begin{align}\label{typicalfrequency}
f_{\rm today} = f_{\rm PT} e^{ - N_{\rm PT} - N_{\rm today}}
%	f_{\text {today }}=f_{*} \frac{a\left(t_{*}\right)}{a_{r}}\left(\frac{g_{* S}^{(0)}}{g_{* S}^{(R)}}\right)^{1 / 3} \frac{T_{\mathrm{CMB}}}{\left[\left(\frac{30}{g_{*}^{(R)} \pi^{2}}\right)\left(\frac{3 H_{r}^{2}}{8 \pi G_{N}}\right)\right]^{1 / 4}} ~. 
\end{align}
where $f_{\rm PT}$ is the GW frequency when it was generated, $N_{\rm PT}$ and $N_{\rm today}$ are the e-folds the Universe expanded from SRFOPT to the end of inflation and from the end of inflation to today, respectively. In the instantaneous reheating scenario, $N_{\rm today}$ can be calculated from today's CMB temperature and the reheating temperature
\bea
e^{- N_{\rm today}} = \frac{T_{\mathrm{CMB}}}{\left[\left(\frac{30}{g_{*}^{(R)} \pi^{2}}\right)\left(\frac{3 H_{r}^{2}}{8 \pi G_{N}}\right)\right]^{1 / 4}} \ ,
\eea
where $H_r$ is the Hubble parameter at the beginning of RD. We further assume the inflation is quasi-de Sitter, and thus we have $H_r\approx H$. 

\begin{figure}[t] 
	\centering 
	\vspace*{0.3cm}
	\includegraphics[width=8.5cm]{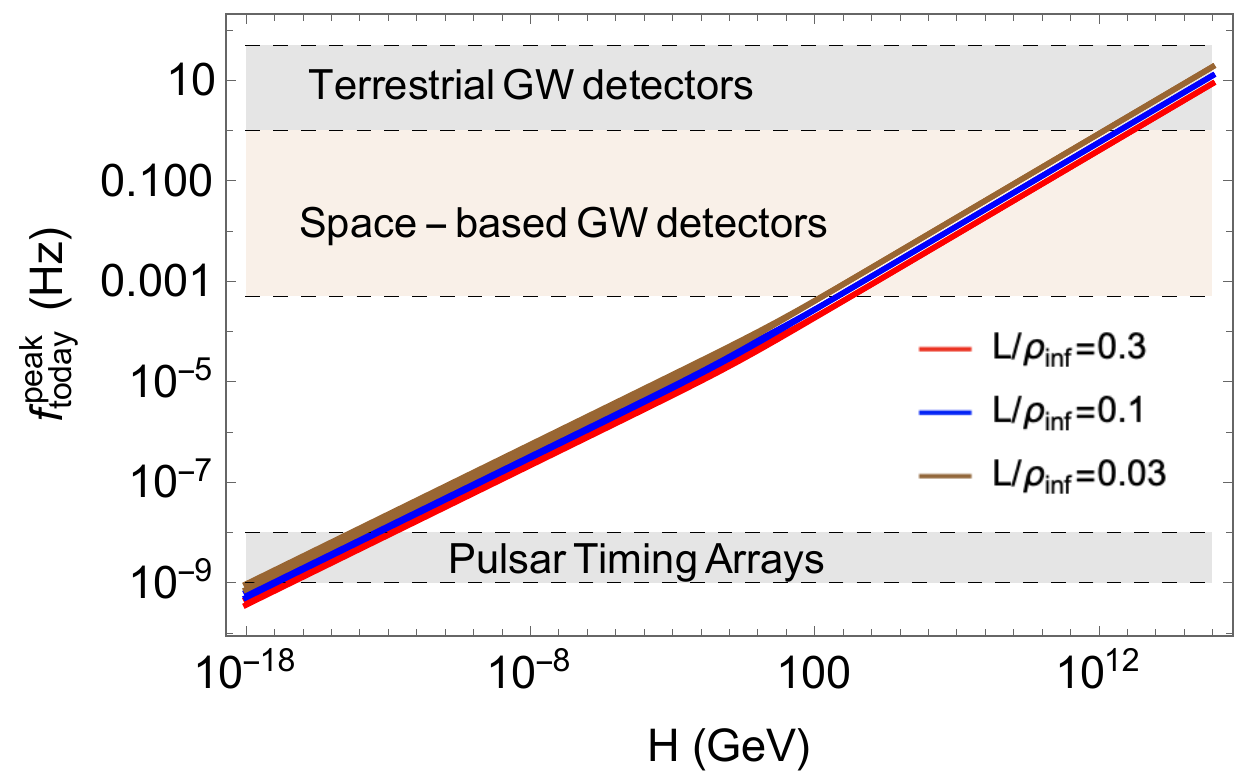}  
	\caption{The frequency of the highest peak of the GW spectrum today produced by SRFOPT during inflation as function of $H$ for various values of $L/\rho_{\rm inf}$. The relic energy density of $\rho_{\rm DM}$ is fixed to today's observed value. The collision parameter ${\cal C}$ in Eq.~(\ref{eq:Yinfty}) is fixed to 0.01. Sensitive regions of future space GW detectors and pulsar timing arrays are also shown by the shaded regions. } \label{fig:fpeak}
\end{figure}

Due to the distortion from inflation, the highest peak of the GW spectrum appears at $f_*^{\rm peak} \approx H/2\pi$~\cite{An:2020fff,An:2022cce}. Together with the relations $L = \gamma_{\rm PT} m_\sigma^4$ and $\rho_{\rm inf} = 3 H^2 M_{\rm pl}^2$, and require $\sigma$ particles constitute all the DM today, we obtain the relation between $f_{\rm today}$ and $H$,
\bea\label{eq:fpeak}
f^{\rm peak}_{\rm today} = \frac{1}{2\pi} \left(\frac{\gamma_{\rm PT}}{\gamma_*}\right)^{1/12} \left( \frac{L}{\rho_{\rm inf}} \right)^{-1/3} \left( \frac{\Omega_{\rm DM} H H_0^2}{Y_\sigma(\infty)} \right)^{1/3},\nn
\eea
where $H_0$ is today's Hubble expansion rate. Assuming $g^{(R)}_{*S}\sim g^{(R)}_*\sim 100$, the peak frequency of today's GW signal is shown in Fig.~\ref{fig:fpeak}. One can see that for high scale inflation scenarios, the corresponding GWs fall in the sensitive region of the space-based GW detectors, such as LISA, Taiji, Tianqin, DECIGO and BBO. For low scale inflation, if $H$ is around $10^{-18} - 10^{-14}$ GeV, the signal may be detected by pulsar time arrays. If $H$ is around $10^{12}-10^{14}$ GeV, the frequency of the corresponding GW signal will be around 1$-$10 Hz, and thus can be detected by future terrestrial GW detectors, such as the Einstein telescope~\cite{Punturo:2010zz} and the Cosmic Explorer~\cite{Reitze:2019iox}.  

%\begin{align}
%	\tilde{f}_{\text {today }}=1.1 \times 10^{11} \mathrm{~Hz}\left(\frac{H}{M_{\mathrm{pl}}}\right)^{1 / 2}\left(\frac{a(t_{*})}{a_{\rm end}}\right)~,
%\end{align}
%where in this case $a_{\rm end}\approx a_r$ is the scale factor when inflation ends. 

\begin{figure*}[t]   
	\includegraphics[width=18cm]{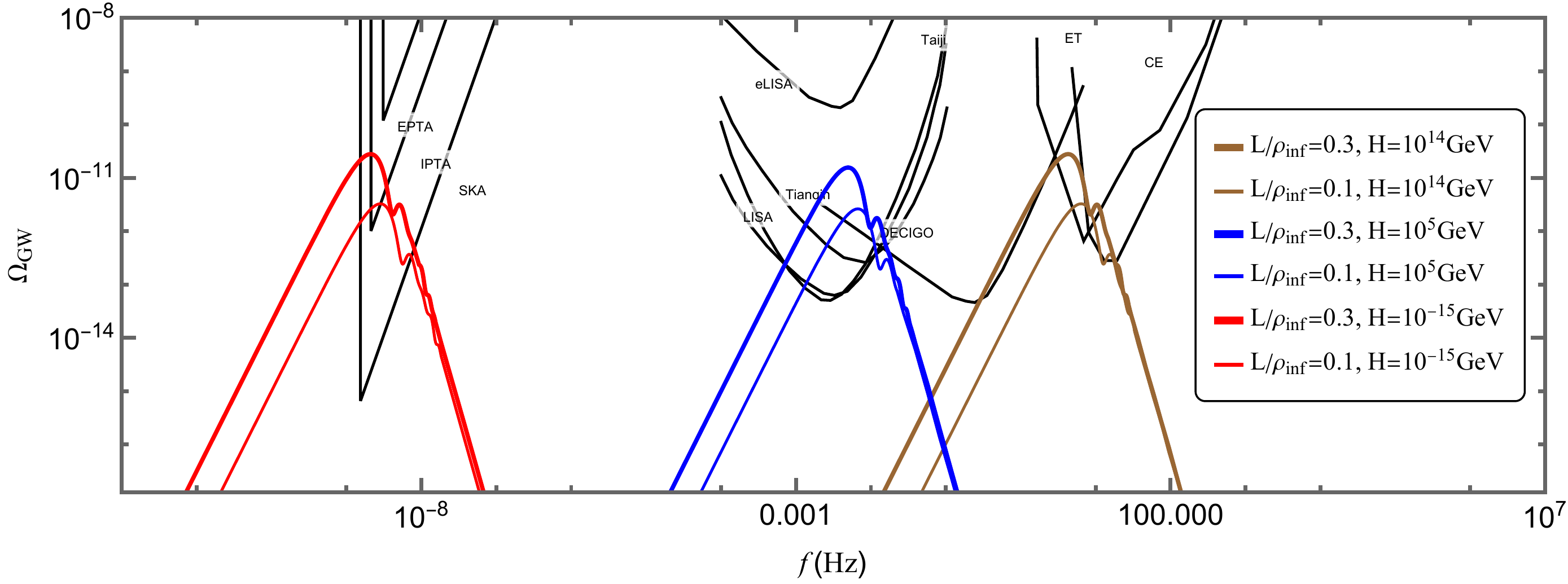}   \\  
	\includegraphics[width=18cm]{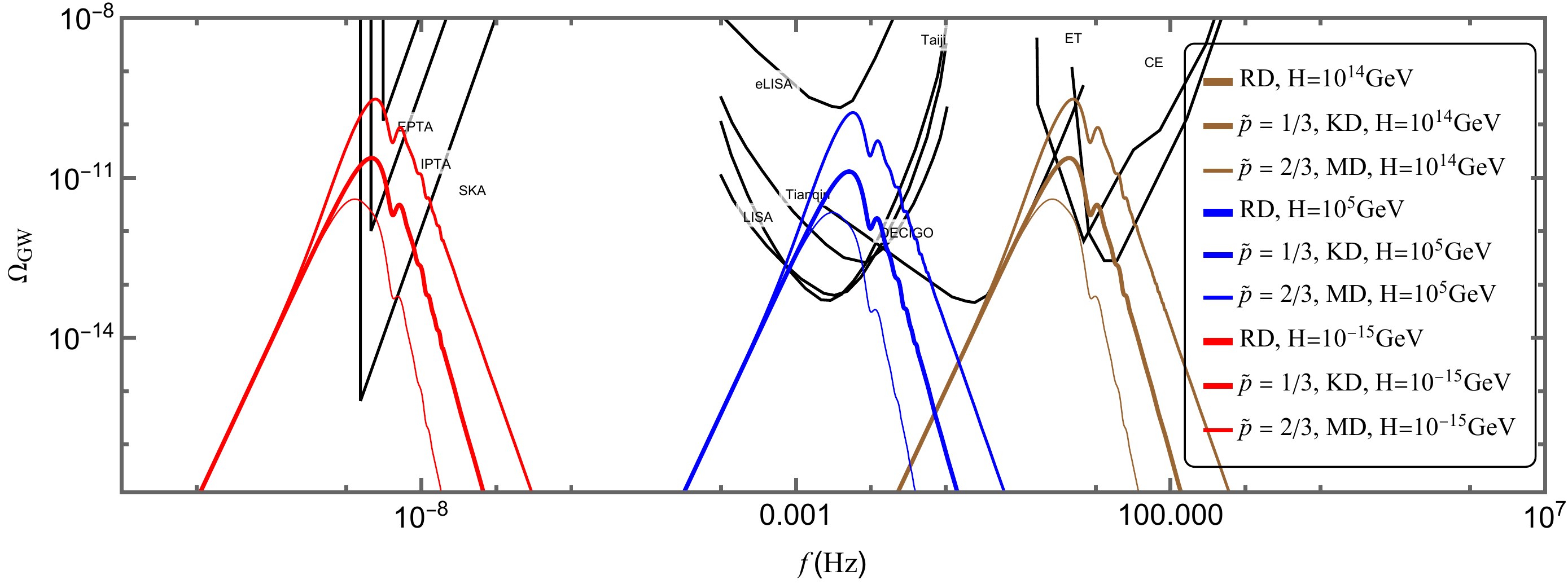}  
	\caption{The upper figure shows the gravitational wave signal associated with the SRFOPT for different superheavy dark matter mass and different e-folding number before the end of inflation. The thickest line corresponds to $L/\rho_{\rm inf}=0.3$ and the think line corresponds to $L/\rho_{\rm inf}=0.1$. The lower figure shows the gravitational wave signature for different universe evolution scenarios. The thickest line denotes the case of instantaneous reheating, the second thickest line denotes the case of kination domination immediately after the end of inflation, the thinest line denotes the case of matter domination immediately after the end of inflation. In the $\tilde p=1/3$ and $\tilde p=2/3$ case, we used $\tau_r/|\tau_{\rm PT}|=10$. In both figures, for the three cases, we take $N_{\rm PT}=18.6, m_\sigma = 170H, H=10^{14} {\rm GeV}$ for the brown curve, $N_{\rm PT}=15, m_\sigma = 5\times 10^6H, H=10^{5} {\rm GeV}$ for the blue curve and  $N_{\rm PT}=7.8, m_\sigma = 1.7\times 10^{16}H, H=10^{-15} {\rm GeV}$ for the red curve. We choose the parameters $N_{\rm PT}$ and $m_\sigma$ such that the SRFOPT can produce just enough $\sigma$ field to explain the DM relic abundance today. The other parameters are chosen to be $\beta = 5H$, $\lambda = -1$, $\gamma_{\rm PT}=0.6$ and $\Lambda = 2 m_\sigma$ when making this plot. The experiments are LISA~\cite{LISA:2017pwj}, eLISA~\cite{eLISA:2013xep}, DECIGO~\cite{Kawamura:2011zz}, UDECIGO~\cite{Yagi:2011wg}, BBO1~\cite{Harry:2006fi}, BBO2~\cite{Corbin:2005ny}, TianQin~\cite{TianQin:2015yph} and Taiji~\cite{Ruan:2018tsw}, ET~\cite{Punturo:2010zz} and CE~\cite{Yang:2022quy}. } \label{GWsignal}
\end{figure*}

The generic features of the GW produced by first-order phase transition during inflation have been studied in \cite{Jiang:2015qor,Wang:2018caj,Li:2020cjj,An:2020fff,An:2022cce}. The IR part of the spectrum has a universal $k^3$ law which is fixed by causality \cite{Caprini:2009fx,Cai:2019cdl}. The spectrum today can be written as 
%\begin{align}
%	\frac{d \rho_{\mathrm{GW}}}{d \ln k}=\frac{a^{4}_r}{a^{4}(\tau)} \mathcal{S}\left(k_{p}\right) \frac{d \rho_{\mathrm{GW}}^{\mathrm{flat}}}{d \ln k_{p}}~,
%\end{align}
%where the form of the shape factor $\mathcal S(k_p)$ can be found in \cite{An:2020fff}. $k_p$ is the physical momentum $k_p\equiv k/a$. The GW spectrum today is given by 
\begin{align}
	\Omega_{\mathrm{GW}}\left(f_{\text {today}}\right)=\Omega_{R}  \mathcal{S}(2\pi f_*)  \bigg( \frac{L}{\rho_{\rm inf}}\bigg) \frac{d \rho_{\mathrm{GW}}^{\mathrm{flat}}}{L \, d \ln f_{p}}~.
\end{align}
where $\Omega_{R}$ is today's abundance of radiation. The GW spectrum in the flat space is given by~\cite{Huber:2008hg}
\begin{align}
	\frac{d \rho_{\mathrm{GW}}^{\mathrm{flat}}}{L d \ln f_{p}}=\kappa^{2}\bigg( \frac{L}{3M_{\rm pl}^2 H^2}\bigg)\left(\frac{H}{\beta}\right)^{2} \Delta\left(2\pi f_{p}\right)~,
\end{align} 
where $\kappa = 1$ since the energy density of the plasma is negligible. The shape function $\Delta(k_{p})$ is fixed by numerical simulation as~\cite{Huber:2008hg}
\begin{align}
	\Delta\left(k_{p}\right)=\tilde{\Delta} \times \frac{3.8 \tilde{k}_{p} k_{p}^{2.8}}{\tilde{k}_{p}^{3.8}+2.8 k_{p}^{3.8}}~.
\end{align}
where $\tilde \Delta\sim 0.077$ and $\tilde k_p\sim 1.44\beta$. The GW spectrums generated in our DM formation model for various choices of $H$, $L/\rho_{\rm inf}$ are shown in Fig.~\ref{GWsignal}. Here, the relic energy density of $\sigma$ particles is fixed to the observed DM relic energy density. The value for $\beta/H$ is fixed to five. For the details of the model parameters, we fixed $\lambda = -1$ and $\Lambda = 2 m_\sigma$, then $m_\sigma$ can be calculated from the latent heat. The collision parameter ${\cal C}$ are calculated in Appendix~\ref{DMReductionRateAppendix}. 

%, and the signal-to-noise ratio for the future BBO1 experiment~\cite{Harry:2006fi}, BBO2 experiment~\cite{Corbin:2005ny}, DECIGO experiment~\cite{Kawamura:2011zz}, UDECIGO experiment~\cite{Yagi:2011wg} is plotted in FIG.~\ref{reachplot}. 
%

\subsection{Intermediate matter/kination domination scenario}
In this section, we discuss the case that inflation is connected to a matter dominated (MD) intermediate stage or a kination dominated (KD) stage before going into radiation domination~\cite{Gouttenoire:2021jhk}. The evolution of the scale factor between inflation and radiation domination is $a(t) = t^{\tilde p}$, $\tilde p=1/3,1/2,2/3$ for kination domination, radiation domination and matter domination, respectively.  The frequency of the highest peak of the GW spectrum does not depend on the physics of the intermediate stages. Since the redshift of GW is completely determined by the expansion of the scale factor $\exp(- N_{\rm today} - N_*)$, which is determined by Eq.~(\ref{eq:rhosigma0}). Thus, we can still use Eq.~(\ref{eq:fpeak}) and Fig.~\ref{fig:fpeak} to estimate the peak frequencies. However, the strengths of the GW signal various with intermediate stages. 

{\bf Matter Domination} In some inflationary models, the inflation will end and the inflaton will oscillate for a while at the end of inflation. This leads to a matter dominated universe at the end of inflation.

{\bf Kination Domination}  Denoting $\tau_r$ as the conformal time at the beginning of RD, $\tau_{\rm PT}$ as the conformal time when the phase transition happens, the e-folding number of the kination domination stage $N_K$ can be computed as 
\begin{align}
	N_{K}=\log \frac{a_{r}}{a_{\text {end }}}=\frac{N_{\rm PT}+\log 2}{2}+\frac{1}{2} \log \frac{\tau_{r}}{\left|\tau_{\rm PT}\right|}~.
\end{align}
The final spectrum for GWs generated in our DM formation model with instantaneous reheating/a kination domination stage between inflation and radiation domination/a matter domination stage between inflation and radiation domination is shown in the lower panel of Fig.~\ref{GWsignal}. In the figure, the brown, purple and red lines correspond to the inflation scale $N_{\rm PT}=18.6, m_\sigma = 170H, H=10^{14} {\rm GeV}$, $N_{\rm PT}=15, m_\sigma = 5\times 10^6H, H=10^{5} {\rm GeV}$ and $N_{\rm PT}=7.8, m_\sigma = 1.7\times 10^{16}H, H=10^{-15} {\rm GeV}$, respectively. As we can see from the figure, the gravitational wave signature with intermediate kination domination has the strongest signal and the gravitational wave signature with intermediate matter domination stage has the weakest signal. During the KD stage, the kination energy density quickly dilutes as $\rho\sim a^{-6}$, whereas the GW energy density still dilutes as $a^{-4}$. The energy density of the radiation today is fixed, so the duration of the KD stage is very short compared with the radiation dominated stage. As a result, the gravitational wave signal has shorter time to decrease. As a result, the observed GW signal is enhanced compared to the instantaneous reheating scenario~\cite{Spokoiny:1993kt,Peebles:1998qn,Gouttenoire:2021jhk}. On the other hand, during the matter dominated stage, the matter energy density dilutes as $\rho \sim a^{-3}$ and the GW energy density dilutes as $a^{-4}$. It will take longer time for the matter energy density to dilute, so the gravitational wave signal will dilute more compared with an intermediate matter dominated stage.

\section{Summary and Discussions}\label{ConclusionSection}

We discuss the possibility that sufficient amount of superheavy DM can be produced during inflation via a SRFOPT. We consider a specific phase transition model where an inflaton $\phi$ and a spectator field $\sigma$ is involved. During the phase transition, the released latent heat can largely contribute to the production of the superheavy DM. After examining the leading processes that reduce $\sigma$ number density, we found the resulting reduction of DM density to be insignificant. As a result, this mechanism can robustly produce the correct DM relic abundance today. We then briefly commented on the potential instability of $\sigma$ particles due to Planckian effects, and suggested an alternative model with local $Z_2$ symmetry for this case. At last, we moved on to the gravitational wave signatures accompanying the phase transition. We showed that the signal frequency falls into the BBO band, with a considerable signal-to-noise ratio, and is thus feasible for future experiments. 

In our minimal model above, the $\sigma$ particle is protected by the restored global $Z_2$ symmetry against decaying, hence they can play the role of DM. However, we comment that (i) symmetry restoration phase transition is not absolutely necessary to generate a $Z_2$ symmetry, and that (ii) a global symmetry may not be sufficient for the stability of $\sigma$. The reason for (i) is that the $Z_2$ symmetry can also be realized as a remainder of a larger symmetry group after breaking, although the corresponding phase transition will still be first-order to generate gravitational waves. As for point (ii), it is widely believed there is no exact global symmetry in quantum gravity \cite{Banks:2010zn,Harlow:2018tng,Harlow:2020bee}. %\xt{cite some} 
Wormhole tunneling processes that violate global charges manifest themselves as Planck-suppressed operators in the low-energy effective Lagrangian \cite{Gilbert:1989nq}. Therefore, quantum gravitational effects may induce the decay of $\sigma$ particles, making them unable to serve as DM. 
%For example, the decay rate induced by a dimension-five operator can be as large as $\Gamma_\sigma\sim \frac{m_\sigma^3}{M_p^2}\sim 10^{10}$GeV for $m_\sigma\simeq 2000H$ and $H\simeq 10^{12}$ GeV. Such a decay rate may be insignificant during inflation ($\Gamma_\sigma\ll H$), but it soon wipes out the $\sigma$ particles in the later stages of the universe.
%
This issue can be solved if we evoke a local $Z_2$ symmetry which arises from the symmetry breaking of a gauge theory. For instance, consider a $U(1)$ gauge theory with two scalars $\sigma,\eta$ that carry charge $+1$ and $+2$, respectively \cite{Krauss:1988zc}. The action reads
\begin{align}
	\nonumber S  = &\int d^4x \sqrt{-g} \Big[-\frac{1}{2}(\partial\phi)^2-|\partial \eta - 2 i g A \eta|^2-W(\phi, |\eta|) \\
	\nonumber& \qquad\quad~- |\partial \sigma - i g A \sigma|^2 - m_\sigma^2 |\sigma|^2\\
	& \qquad\quad~ -\frac{1}{4}F_{\mu\nu}F^{\mu\nu}\Big].
\end{align}
Similar to (\ref{Upotential}), $W(\phi,|\eta|)$ is a potential providing a slow-rolling inflaton $\phi$ that dynamically generates a VEV for the $\eta$ field. This leads to a symmetry-breaking phase transition with bubble collision and $\sigma$ particle production. Now due to the charge difference, the $U(1)$ gauge symmetry is broken down to a \textit{local} $Z_2$ symmetry:
\begin{align}
	\nonumber\sigma&\to e^{i\alpha(x)}\sigma=\pm \sigma,\\
	\nonumber\eta&\to e^{2i\alpha(x)}\eta=\eta,\\
	A&\to A+\frac{1}{g}\partial\alpha(x),\quad\text{with }\alpha(x)=0,\pi~.
\end{align}
Since the local $Z_2$ is essentially a gauge redundancy, it must still be respected in quantum gravity. In effect, $Z_2$-violating operators that leads to $\sigma$ decay are strictly forbidden in the effective Lagrangian. Thus the $\sigma$ particles thermally produced in the plasma of phase transition can serve as stable DM.
In the case where $\sigma$ is the lightest massive particle, $i.e.$, $H\ll m_\sigma\ll m_\eta, m_A$, the latent heat of phase transition is almost entirely stored in the $\sigma$ sector. The calculations of relic abundance and GW production are similar to our original model and will be skipped here for simplicity.

\begin{acknowledgments} 
We would like to thank Kunfeng Lyu, Toshifumi Noumi, Pablo Soler, Toshiaki Takeuchi, Ke-Pan Xie and Ziwei Wang for valuable discussions. HA is supported in part by the National Key R\&D Program of China under Grant No. 2021YFC2203100 and 2017YFA0402204, the NSFC under Grant No. 11975134, and the Tsinghua University Initiative Scientific Research Program. XT is supported by National Key R\&D Program of China (2021YFC2203100). The work of SZ is supported by in part by JSPS KAKENHI Grant Number 21F21026.
\end{acknowledgments}

\appendix
\section{Estimation of Dark Matter Reduction Rate}\label{DMReductionRateAppendix} 
\begin{figure}[t] 
	\centering 
	\includegraphics[width=8cm]{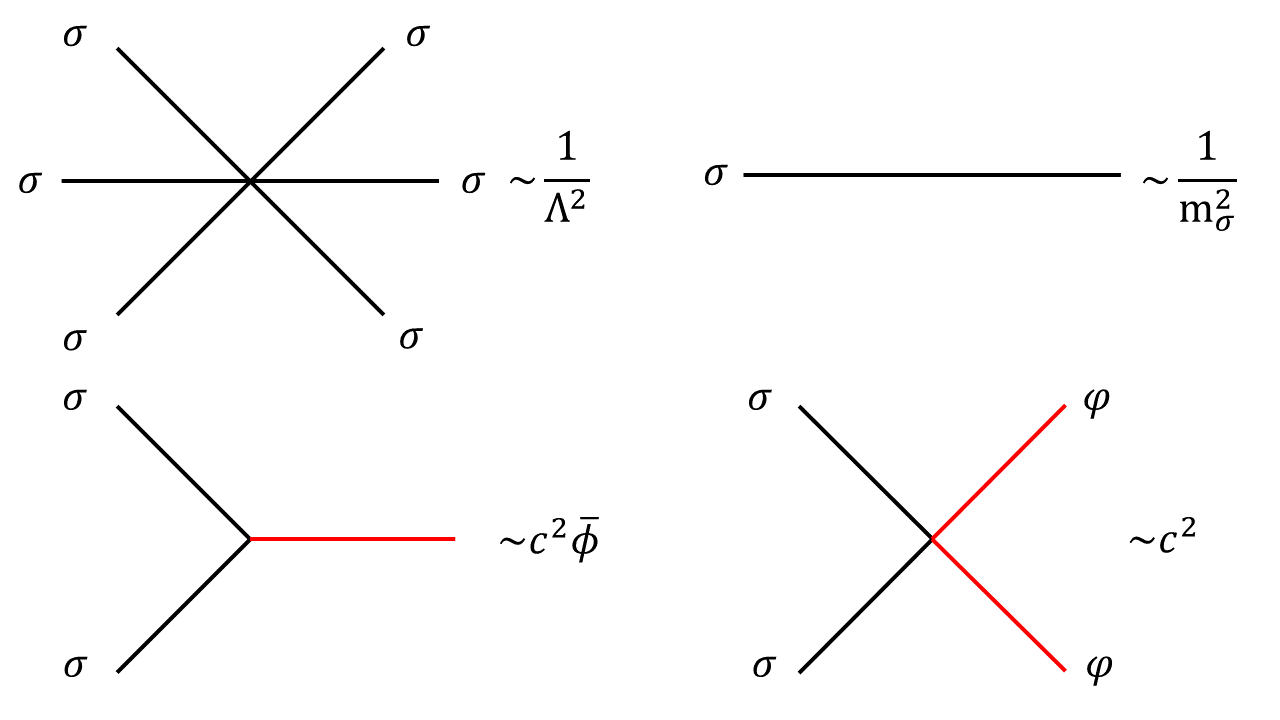}  
	\caption{The magnitude estimation of different elements that will contribute to the scattering amplitude in the model. We can use it to estimate the dominant processes that contribute to DM reduction.} \label{magnitude}
\end{figure}

In this appendix, we estimate the DM reduction rate. In order to find out the dominant process that contributes to the superheavy DM decay, it is useful to first examine how large each vertex will contribute to certain Feynman diagrams. We listed them in Fig.~\ref{magnitude}.

\subsubsection{Processes due to $\sigma$ self-interaction}
The collision term $C_{\sigma\sigma\sigma\sigma\rightarrow \sigma\sigma}$ relevant for this channel is 
\begin{align}\nonumber
	&C_{\sigma\sigma\sigma\sigma\rightarrow\sigma\sigma} \\\nonumber
	=&~ \prod_{i=1}^{6} \int \frac{d^3\mathbf p_i}{(2\pi)^3 2 E_i} (2\pi)^4 \delta^3(\mathbf p_1 + \mathbf p_2 +\mathbf p_3+\mathbf p_4-\mathbf p_5-\mathbf p_6) \\\nonumber
	& ~~~~~\times\delta (E_1+E_2+E_3+E_4-E_5-E_6)\\\nonumber
	&  ~~~~~\times\left[|\mathcal M_{\rm 2\to 4}|^2 f^2(1+f)^4 - |\mathcal M_{\rm 4\to 2}|^2 f^4(1+f)^2\right]~.
\end{align}
where $f$ is given as a Maxwell distribution as in \eqref{MBdistribution}, now with a time-dependent temperature $T(t)$. $\mathcal M_{\rm 2\to 4}$ and $\mathcal M_{\rm 4\to 2}$ denotes the scattering amplitudes for the $2\to4$ process and $4\to2$ process, respectively. As mentioned before, these two processes balance each other until the particles become non-relativistic at $\gamma_*\simeq 0.01$. In this regime, $f\ll 1$, so we can approximate $1+f\sim 1$. So the collision term simplifies to
\begin{align}\nonumber
	&C_{\sigma\sigma\sigma\sigma\rightarrow\sigma\sigma} \\ \nonumber
	=&- \prod_{i=1}^{6} \int \frac{d^3\mathbf p_i}{(2\pi)^3 2 E_i} (2\pi)^4 \delta^3(\mathbf p_1 + \mathbf p_2 +\mathbf p_3+\mathbf p_4-\mathbf p_5-\mathbf p_6) \\
	&~~~~~~~\times \delta (E_1+E_2+E_3+E_4-E_5-E_6)|\mathcal M_{\rm 4\to 2}|^2 f^4 ~.
\end{align} 
The total amplitude consists of three parts (see FIG.~\ref{2to2}), $\mathcal M_{\rm 4\to 2} = \mathcal M_{\rm contact} + \mathcal M_{\rm exchange\,\, 1}+\mathcal M_{\rm exchange\,\, 2}$, whose sizes in the non-relativistic limit read
\begin{align}
	& \mathcal M_{\rm contact} =    \frac{90}{\Lambda^2}~,  \\
	& \mathcal M_{\rm exchange\,\, 1}   =- \frac{18\lambda^2}{m_\sigma^2}~, \\
	& \mathcal M_{\rm exchange\,\, 2}   = \frac{54\lambda^2}{m_\sigma^2}~.
\end{align}
We have assumed that the four particles are non-relativistic at the initial state, so $E(\mathbf p_i) \simeq m_\sigma$ for $i=1,\ldots,4$.  
The corresponding term $C_{\sigma\sigma\sigma\sigma\rightarrow \sigma\sigma}$ in the Boltzmann equation is thus
\begin{align}\nonumber
	 C_{\sigma\sigma\sigma\sigma\rightarrow\sigma\sigma}   & = -\frac{1}{4!2!}  \frac{1}{256 m_\sigma^4} \frac{\sqrt{3}}{ \pi}   \bigg|\frac{90}{\Lambda^2}- \frac{18\lambda^2}{m_\sigma^2}+\frac{54\lambda^2}{m_\sigma^2}\bigg|^2 n_\sigma^4~ \\ \label{Boltzmann6sigma}
%	&= - 5\times 10^{-6} \gamma_*^4  m_\sigma^8 \left(\frac{90}{\Lambda ^2}+\frac{36 \lambda ^2}{ m_\sigma^2}\right)^2 \nn
	&\equiv-{\cal C} \gamma_*^4 m_\sigma^4 \times \frac{n_\sigma^4}{n_\sigma^4(t_*)}~,
\end{align}
where $n_\sigma(t)=\int d^3\mathbf p/(2\pi)^3 f(E(p),T(t))$ is the time-dependent particle number density. The collision parameter ${\cal C}$ defined here is used in Eq.~(\ref{eq:dT}).

\subsubsection{Processes involving interactions with the inflaton} 
The other processes that might contribute to the decrease of the number density of the superheavy DM are: (a). $\sigma\sigma\rightarrow\varphi\varphi$, where two DM particles decay into two inflatons and (b). $\sigma\sigma\sigma\rightarrow\sigma\varphi$, where three DM particles decay into one DM particle and one inflaton. Following the same procedure, we can obtain the collision terms $C_{\sigma\sigma\rightarrow\varphi\varphi}$ and $C_{\sigma\sigma\sigma\rightarrow\sigma\varphi}$. The relevant Feynman diagrams are depicted in FIG.~\ref{2to2}.  
\begin{align}\label{Boltzmaninflaton}
	C_{\sigma\sigma\rightarrow\varphi\varphi} & = -\frac{1}{32\pi m_\sigma^2} \bigg| -c^2-\frac{c^4\bar\phi^2}{m_\sigma^2} \bigg|^2 n_\sigma^2 ~, \\
	C_{\sigma\sigma\sigma\rightarrow\sigma\varphi} & = \frac{1}{3!} \frac{1}{72\pi m_\sigma^3}   \bigg(  - \bigg|\frac{12\lambda c^2\bar \phi}{m_\sigma^2}\bigg|^2 n_\sigma^3 \bigg)~.\label{Boltzmann3sigma1sigma1phi}
\end{align}   

\subsubsection{Number density evolution}
\begin{figure}[t] 
	\centering 
	\vspace*{0.3cm}
	\includegraphics[width=8.5cm]{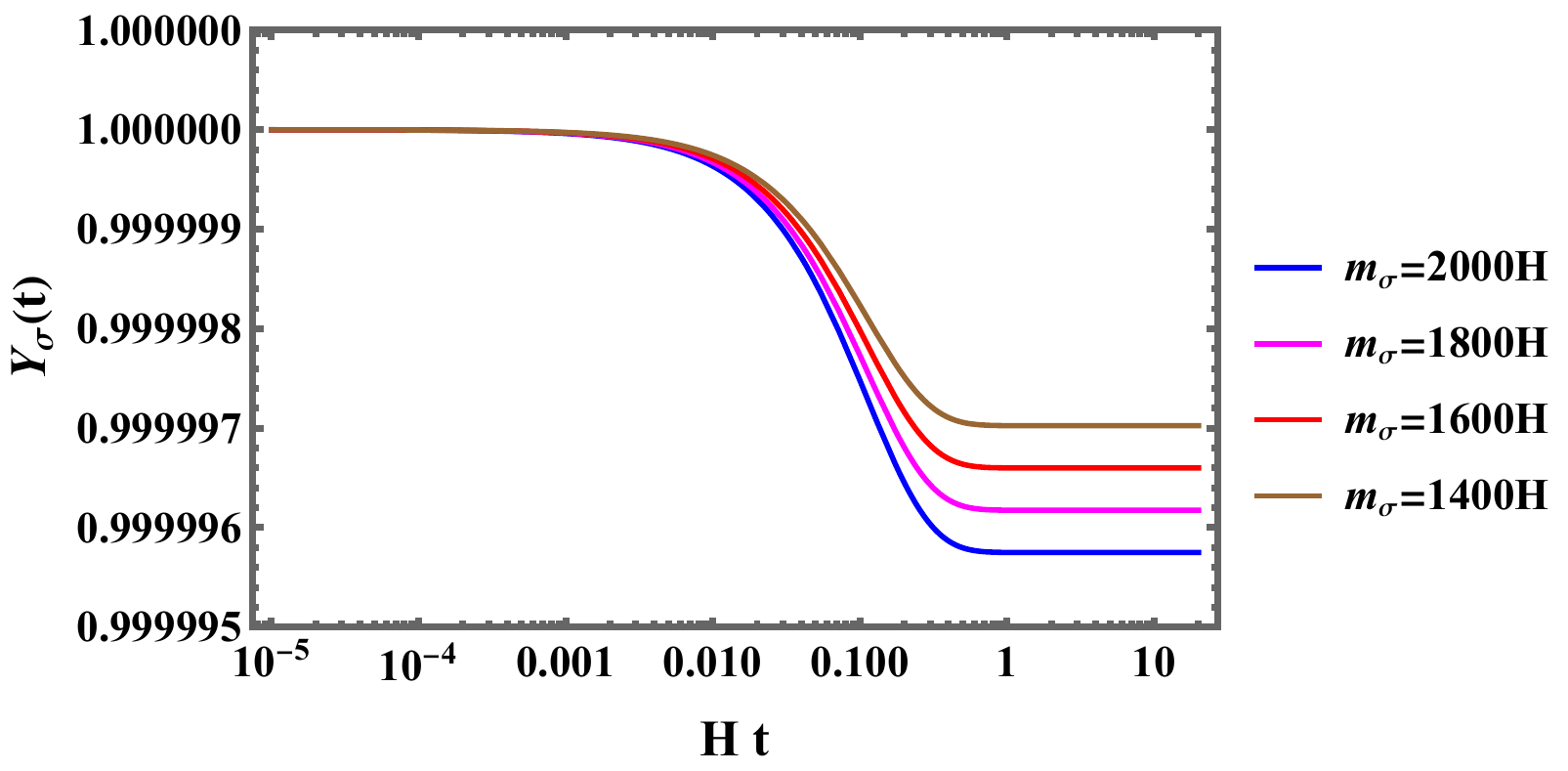}  
	\caption{This figure shows the evolution of DM comoving particle number density after considering its interactions. The parameters are chosen as $\lambda = -1$, $\gamma_*=0.01$ and $\Lambda = 2 m_\sigma$ when making this plot.} \label{relic}
\end{figure}

As stated in the main text, the evolution of $n_\sigma(t)$ can be solved by combining (\ref{thermalization}) with (\ref{Boltzmann6sigma},\ref{Boltzmaninflaton},\ref{Boltzmann3sigma1sigma1phi}), with an initial condition $n_\sigma(t_*)=\gamma_* m_\sigma^3/2$. While the full numeric solution is always attainable, we can already extract some useful information by comparing the magnitude of $C_{\sigma\sigma\sigma\sigma\rightarrow\sigma\sigma}$, $C_{\sigma\sigma\rightarrow\varphi\varphi}$ and $C_{\sigma\sigma\sigma\rightarrow\sigma\varphi}$. Inserting a set of typical parameters at $t=t_*$,
\begin{align}
	\lambda = -1~,\quad \Lambda \sim 2 m_\sigma\sim 2000H~, \quad\gamma_*\sim 0.01~,\label{suitableChoice}
\end{align}
we arrive at %\xt{Check numerics}
\begin{align}
	\nonumber C_{\sigma\sigma\sigma\sigma\rightarrow\sigma\sigma} & \sim - 9.6\times10^{-11} \, m_\sigma^4 ~, \\
	\nonumber C_{\sigma\sigma\rightarrow\varphi\varphi} & \sim - 1.8\times 10^{-14} \, m_\sigma^4~, \\
	C_{\sigma\sigma\sigma\rightarrow\sigma\varphi} & \sim - 2.2\times 10^{-12} \, m_\sigma^4~.
\end{align} 
Clearly, the first term dominates over the other two. Considering the first term only, the Boltzmann equation (\ref{thermalization}) can be solved analytically:
\begin{align}  
	Y_\sigma(t)=\frac{ H^3}{\bigg( (\frac{\mathcal C H^8}{3\times16 m_\sigma^{8}} + \frac{H^9}{n_\sigma(t_*)^3})- \frac{a(t_*)^9}{a(t)^9} \frac{\mathcal C H^8}{3\times16 m_\sigma^{8}} \bigg)^{1/3}n_\sigma(t_*)}~.
\end{align} 
The full behavior of $Y_\sigma(t)$ is shown in FIG.~\ref{relic}%\xt{Re-draw}
. It can be seen that the various annihilation processes reduce the $\sigma$ particle density at a constant rate initially, and become suppressed as the universe expands. Within one e-fold, the comoving particle number density is frozen out to a value not far below the initial one. As a result, we conclude that with suitable parameter choices (such as (\ref{suitableChoice})), the comoving density of DM is not significantly reduced, and the single-field approximation for computing relic abundance in Sect.~\ref{ModelSection} is justified. 

%---------   References   ---------%


\begin{thebibliography}{99}
%\cite{Griest:1989wd}
\bibitem{Griest:1989wd}
K.~Griest and M.~Kamionkowski,
``Unitarity Limits on the Mass and Radius of Dark Matter Particles,''
Phys. Rev. Lett. \textbf{64} (1990), 615
%doi:10.1103/PhysRevLett.64.615
%733 citations counted in INSPIRE as of 04 Apr 2022

%\cite{McDonald:2001vt}
\bibitem{McDonald:2001vt}
J.~McDonald,
``Thermally generated gauge singlet scalars as selfinteracting dark matter,''
Phys. Rev. Lett. \textbf{88} (2002), 091304
%doi:10.1103/PhysRevLett.88.091304
[arXiv:hep-ph/0106249 [hep-ph]].
%351 citations counted in INSPIRE as of 05 Apr 2022

%\cite{Hall:2009bx}
\bibitem{Hall:2009bx}
L.~J.~Hall, K.~Jedamzik, J.~March-Russell and S.~M.~West,
``Freeze-In Production of FIMP Dark Matter,''
JHEP \textbf{03} (2010), 080
%doi:10.1007/JHEP03(2010)080
[arXiv:0911.1120 [hep-ph]].
%864 citations counted in INSPIRE as of 05 Apr 2022

%\cite{Bernal:2017kxu}
\bibitem{Bernal:2017kxu}
N.~Bernal, M.~Heikinheimo, T.~Tenkanen, K.~Tuominen and V.~Vaskonen,
``The Dawn of FIMP Dark Matter: A Review of Models and Constraints,''
Int. J. Mod. Phys. A \textbf{32} (2017) no.27, 1730023
%doi:10.1142/S0217751X1730023X
[arXiv:1706.07442 [hep-ph]].
%322 citations counted in INSPIRE as of 05 Apr 2022

%\cite{Goudelis:2018xqi}
\bibitem{Goudelis:2018xqi}
A.~Goudelis, K.~A.~Mohan and D.~Sengupta,
``Clockworking FIMPs,''
JHEP \textbf{10} (2018), 014
%doi:10.1007/JHEP10(2018)014
[arXiv:1807.06642 [hep-ph]].
%29 citations counted in INSPIRE as of 05 Apr 2022

%\cite{Chowdhury:2018tzw}
\bibitem{Chowdhury:2018tzw}
D.~Chowdhury, E.~Dudas, M.~Dutra and Y.~Mambrini,
``Moduli Portal Dark Matter,''
Phys. Rev. D \textbf{99} (2019) no.9, 095028
%doi:10.1103/PhysRevD.99.095028
[arXiv:1811.01947 [hep-ph]].
%37 citations counted in INSPIRE as of 05 Apr 2022

%\cite{Bhattacharyya:2018evo}
\bibitem{Bhattacharyya:2018evo}
G.~Bhattacharyya, M.~Dutra, Y.~Mambrini and M.~Pierre,
``Freezing-in dark matter through a heavy invisible Z',''
Phys. Rev. D \textbf{98} (2018) no.3, 035038
%doi:10.1103/PhysRevD.98.035038
[arXiv:1806.00016 [hep-ph]].
%42 citations counted in INSPIRE as of 05 Apr 2022

%\cite{Bernal:2018qlk}
\bibitem{Bernal:2018qlk}
N.~Bernal, M.~Dutra, Y.~Mambrini, K.~Olive, M.~Peloso and M.~Pierre,
``Spin-2 Portal Dark Matter,''
Phys. Rev. D \textbf{97} (2018) no.11, 115020
%doi:10.1103/PhysRevD.97.115020
[arXiv:1803.01866 [hep-ph]].
%68 citations counted in INSPIRE as of 05 Apr 2022

%\cite{Chung:1998rq}
\bibitem{Chung:1998rq}
D.~J.~H.~Chung, E.~W.~Kolb and A.~Riotto,
``Production of massive particles during reheating,''
Phys. Rev. D \textbf{60} (1999), 063504
%doi:10.1103/PhysRevD.60.063504
[arXiv:hep-ph/9809453 [hep-ph]].
%397 citations counted in INSPIRE as of 05 Apr 2022

%\cite{Chung:1998ua}
\bibitem{Chung:1998ua}
D.~J.~H.~Chung, E.~W.~Kolb and A.~Riotto,
``Nonthermal supermassive dark matter,''
Phys. Rev. Lett. \textbf{81} (1998), 4048-4051
%doi:10.1103/PhysRevLett.81.4048
[arXiv:hep-ph/9805473 [hep-ph]].
%271 citations counted in INSPIRE as of 05 Apr 2022

%\cite{Giudice:2000ex}
\bibitem{Giudice:2000ex}
G.~F.~Giudice, E.~W.~Kolb and A.~Riotto,
``Largest temperature of the radiation era and its cosmological implications,''
Phys. Rev. D \textbf{64} (2001), 023508
%doi:10.1103/PhysRevD.64.023508
[arXiv:hep-ph/0005123 [hep-ph]].
%519 citations counted in INSPIRE as of 05 Apr 2022

%\cite{Garny:2015sjg}
\bibitem{Garny:2015sjg}
M.~Garny, M.~Sandora and M.~S.~Sloth,
``Planckian Interacting Massive Particles as Dark Matter,''
Phys. Rev. Lett. \textbf{116} (2016) no.10, 101302
%doi:10.1103/PhysRevLett.116.101302
[arXiv:1511.03278 [hep-ph]].
%100 citations counted in INSPIRE as of 04 Apr 2022

%\cite{Garny:2017kha}
\bibitem{Garny:2017kha}
M.~Garny, A.~Palessandro, M.~Sandora and M.~S.~Sloth,
``Theory and Phenomenology of Planckian Interacting Massive Particles as Dark Matter,''
JCAP \textbf{02} (2018), 027
%doi:10.1088/1475-7516/2018/02/027
[arXiv:1709.09688 [hep-ph]].
%74 citations counted in INSPIRE as of 04 Apr 2022

%\cite{Hashiba:2018iff}
\bibitem{Hashiba:2018iff}
S.~Hashiba and J.~Yokoyama,
``Gravitational reheating through conformally coupled superheavy scalar particles,''
JCAP \textbf{01} (2019), 028
%doi:10.1088/1475-7516/2019/01/028
[arXiv:1809.05410 [gr-qc]].
%39 citations counted in INSPIRE as of 04 Apr 2022

%\cite{Hashiba:2018tbu}
\bibitem{Hashiba:2018tbu}
S.~Hashiba and J.~Yokoyama,
``Gravitational particle creation for dark matter and reheating,''
Phys. Rev. D \textbf{99} (2019) no.4, 043008
%doi:10.1103/PhysRevD.99.043008
[arXiv:1812.10032 [hep-ph]].
%46 citations counted in INSPIRE as of 04 Apr 2022

%\cite{Haro:2018zdb}
\bibitem{Haro:2018zdb}
J.~Haro, W.~Yang and S.~Pan,
``Reheating in quintessential inflation via gravitational production of heavy massive particles: A detailed analysis,''
JCAP \textbf{01} (2019), 023
%doi:10.1088/1475-7516/2019/01/023
[arXiv:1811.07371 [gr-qc]].
%36 citations counted in INSPIRE as of 04 Apr 2022

%\cite{Feng:2010gw}
\bibitem{Feng:2010gw}
J.~L.~Feng,
``Dark Matter Candidates from Particle Physics and Methods of Detection,''
Ann. Rev. Astron. Astrophys. \textbf{48} (2010), 495-545
%doi:10.1146/annurev-astro-082708-101659
[arXiv:1003.0904 [astro-ph.CO]].
%1193 citations counted in INSPIRE as of 06 May 2022

%\cite{Ema:2019yrd}
\bibitem{Ema:2019yrd}
Y.~Ema, K.~Nakayama and Y.~Tang,
``Production of purely gravitational dark matter: the case of fermion and vector boson,''
JHEP \textbf{07} (2019), 060
%doi:10.1007/JHEP07(2019)060
[arXiv:1903.10973 [hep-ph]].
%66 citations counted in INSPIRE as of 06 May 2022

%\cite{Ahmed:2020fhc}
\bibitem{Ahmed:2020fhc}
A.~Ahmed, B.~Grzadkowski and A.~Socha,
``Gravitational production of vector dark matter,''
JHEP \textbf{08} (2020), 059
%doi:10.1007/JHEP08(2020)059
[arXiv:2005.01766 [hep-ph]].
%43 citations counted in INSPIRE as of 06 May 2022 

%\cite{Gross:2020zam}
\bibitem{Gross:2020zam}
C.~Gross, S.~Karamitsos, G.~Landini and A.~Strumia,
``Gravitational Vector Dark Matter,''
JHEP \textbf{03} (2021), 174
%doi:10.1007/JHEP03(2021)174
[arXiv:2012.12087 [hep-ph]].
%12 citations counted in INSPIRE as of 07 May 2022

%\cite{Ahmed:2021fvt}
\bibitem{Ahmed:2021fvt}
A.~Ahmed, B.~Grzadkowski and A.~Socha,
``Implications of time-dependent inflaton decay on reheating and dark matter production,''
[arXiv:2111.06065 [hep-ph]].
%6 citations counted in INSPIRE as of 07 May 2022

%\cite{Xue:2021jyj}
\bibitem{Xue:2021jyj}
S.~S.~Xue,
``Massive particle pair production and oscillation in Friedman Universe: its consequence on inflation,''
[arXiv:2112.09661 [gr-qc]].
%0 citations counted in INSPIRE as of 07 May 2022

%\cite{Ema:2021fdz}
\bibitem{Ema:2021fdz}
Y.~Ema, K.~Mukaida and K.~Nakayama,
``Scalar field couplings to quadratic curvature and decay into gravitons,''
[arXiv:2112.12774 [hep-ph]].
%1 citations counted in INSPIRE as of 07 May 2022

%\cite{Clery:2022wib}
\bibitem{Clery:2022wib}
S.~Clery, Y.~Mambrini, K.~A.~Olive, A.~Shkerin and S.~Verner,
``Gravitational Portals with Non-Minimal Couplings,''
[arXiv:2203.02004 [hep-ph]].
%1 citations counted in INSPIRE as of 07 May 2022

%\cite{Wang:2022ojc}
\bibitem{Wang:2022ojc}
Q.~Y.~Wang, Y.~Tang and Y.~L.~Wu,
``Dark Matter Production in Weyl $R^2$ Inflation,''
[arXiv:2203.15452 [hep-ph]].
%0 citations counted in INSPIRE as of 07 May 2022

%\cite{Carney:2022gse}
\bibitem{Carney:2022gse}
D.~Carney, N.~Raj, Y.~Bai, J.~Berger, C.~Blanco, J.~Bramante, C.~Cappiello, M.~Dutra, R.~Ebadi and K.~Engel, \textit{et al.}
``Snowmass2021 Cosmic Frontier White Paper: Ultraheavy particle dark matter,''
[arXiv:2203.06508 [hep-ph]].
%0 citations counted in INSPIRE as of 04 Apr 2022

%\cite{Parker:1969au}
\bibitem{Parker:1969au}
L.~Parker,
``Quantized fields and particle creation in expanding universes. 1.,''
Phys. Rev. \textbf{183} (1969), 1057-1068
%doi:10.1103/PhysRev.183.1057
%1054 citations counted in INSPIRE as of 06 Apr 2022

%\cite{Ford:1986sy}
\bibitem{Ford:1986sy}
L.~H.~Ford,
``Gravitational Particle Creation and Inflation,''
Phys. Rev. D \textbf{35} (1987), 2955
%doi:10.1103/PhysRevD.35.2955
%472 citations counted in INSPIRE as of 06 Apr 2022

%\cite{Li:2019ves}
\bibitem{Li:2019ves}
L.~Li, T.~Nakama, C.~M.~Sou, Y.~Wang and S.~Zhou,
``Gravitational Production of Superheavy Dark Matter and Associated Cosmological Signatures,''
JHEP \textbf{07} (2019), 067
%doi:10.1007/JHEP07(2019)067
[arXiv:1903.08842 [astro-ph.CO]].
%38 citations counted in INSPIRE as of 06 Apr 2022

%\cite{Li:2020xwr}
\bibitem{Li:2020xwr}
L.~Li, S.~Lu, Y.~Wang and S.~Zhou,
``Cosmological Signatures of Superheavy Dark Matter,''
JHEP \textbf{07} (2020), 231
%doi:10.1007/JHEP07(2020)231
[arXiv:2002.01131 [hep-ph]].
%20 citations counted in INSPIRE as of 06 Apr 2022

%\cite{Ling:2021zlj}
\bibitem{Ling:2021zlj}
S.~Ling and A.~J.~Long,
``Superheavy scalar dark matter from gravitational particle production in $\alpha$-attractor models of inflation,''
Phys. Rev. D \textbf{103} (2021) no.10, 103532
%doi:10.1103/PhysRevD.103.103532
[arXiv:2101.11621 [astro-ph.CO]].
%9 citations counted in INSPIRE as of 04 Apr 2022

%\cite{Sou:2021juh}
\bibitem{Sou:2021juh}
C.~M.~Sou, X.~Tong and Y.~Wang,
``Chemical-potential-assisted particle production in FRW spacetimes,''
JHEP \textbf{06} (2021), 129
%doi:10.1007/JHEP06(2021)129
[arXiv:2104.08772 [hep-th]].
%7 citations counted in INSPIRE as of 06 Apr 2022

%\cite{Hashiba:2020rsi}
\bibitem{Hashiba:2020rsi}
S.~Hashiba, Y.~Yamada and J.~Yokoyama,
``Particle production induced by vacuum decay in real time dynamics,''
Phys. Rev. D \textbf{103} (2021) no.4, 045006
%doi:10.1103/PhysRevD.103.045006
[arXiv:2006.10986 [hep-th]].
%5 citations counted in INSPIRE as of 07 May 2022

%\cite{Hashiba:2021npn}
\bibitem{Hashiba:2021npn}
S.~Hashiba and Y.~Yamada,
``Stokes phenomenon and gravitational particle production \textemdash{} How to evaluate it in practice,''
JCAP \textbf{05} (2021), 022
%doi:10.1088/1475-7516/2021/05/022
[arXiv:2101.07634 [hep-th]].
%11 citations counted in INSPIRE as of 07 May 2022

%\cite{Yamada:2021kqw}
\bibitem{Yamada:2021kqw}
Y.~Yamada,
``Superadiabatic basis in cosmological particle production: application to preheating,''
JCAP \textbf{09} (2021), 009
%doi:10.1088/1475-7516/2021/09/009
[arXiv:2106.06111 [hep-th]].
%1 citations counted in INSPIRE as of 07 May 2022

%\cite{Basso:2021whd}
\bibitem{Basso:2021whd}
E.~E.~Basso and D.~J.~H.~Chung,
``Computation of gravitational particle production using adiabatic invariants,''
JHEP \textbf{11} (2021), 146
%doi:10.1007/JHEP11(2021)146
[arXiv:2108.01653 [hep-ph]].
%3 citations counted in INSPIRE as of 07 May 2022

%\cite{Ford:2021syk}
\bibitem{Ford:2021syk}
L.~H.~Ford,
``Cosmological particle production: a review,''
Rept. Prog. Phys. \textbf{84} (2021) no.11, 116901
%doi:10.1088/1361-6633/ac1b23
[arXiv:2112.02444 [gr-qc]].
%4 citations counted in INSPIRE as of 04 Apr 2022

%\cite{Lyth:1996im}
\bibitem{Lyth:1996im}
D.~H.~Lyth,
``What would we learn by detecting a gravitational wave signal in the cosmic microwave background anisotropy?,''
Phys. Rev. Lett. \textbf{78} (1997), 1861-1863
%doi:10.1103/PhysRevLett.78.1861
[arXiv:hep-ph/9606387 [hep-ph]].
%796 citations counted in INSPIRE as of 06 May 2022

%\cite{Babichev:2020xeg}
\bibitem{Babichev:2020xeg}
E.~Babichev, D.~Gorbunov and S.~Ramazanov,
``Gravitational misalignment mechanism of Dark Matter production,''
JCAP \textbf{08} (2020), 047
%doi:10.1088/1475-7516/2020/08/047
[arXiv:2004.03410 [hep-ph]].
%8 citations counted in INSPIRE as of 19 Apr 2022

%\cite{Laulumaa:2020pqi}
\bibitem{Laulumaa:2020pqi}
L.~Laulumaa, T.~Markkanen and S.~Nurmi,
``Primordial dark matter from curvature induced symmetry breaking,''
JCAP \textbf{08} (2020), 002
%doi:10.1088/1475-7516/2020/08/002
[arXiv:2005.04061 [astro-ph.CO]].
%6 citations counted in INSPIRE as of 19 Apr 2022

%\cite{Karam:2020rpa}
\bibitem{Karam:2020rpa}
A.~Karam, M.~Raidal and E.~Tomberg,
``Gravitational dark matter production in Palatini preheating,''
JCAP \textbf{03} (2021), 064
%doi:10.1088/1475-7516/2021/03/064
[arXiv:2007.03484 [astro-ph.CO]].
%23 citations counted in INSPIRE as of 19 Apr 2022

%\cite{Borah:2020ljr}
\bibitem{Borah:2020ljr}
D.~Borah, S.~Jyoti Das and A.~K.~Saha,
``Gravitational origin of dark matter and Majorana neutrino mass with non-minimal quartic inflation,''
Phys. Dark Univ. \textbf{33} (2021), 100858
%doi:10.1016/j.dark.2021.100858
[arXiv:2011.02489 [hep-ph]].
%0 citations counted in INSPIRE as of 19 Apr 2022

%\cite{Ramazanov:2021eya}
\bibitem{Ramazanov:2021eya}
S.~Ramazanov, E.~Babichev, D.~Gorbunov and A.~Vikman,
``Beyond freeze-in: Dark matter via inverse phase transition and gravitational wave signal,''
Phys. Rev. D \textbf{105} (2022) no.6, 063530
%doi:10.1103/PhysRevD.105.063530
[arXiv:2104.13722 [hep-ph]].
%7 citations counted in INSPIRE as of 19 Apr 2022

%\cite{Jiang:2015qor}
\bibitem{Jiang:2015qor}
H.~Jiang, T.~Liu, S.~Sun and Y.~Wang,
%``Echoes of Inflationary First-Order Phase Transitions in the CMB,''
Phys. Lett. B \textbf{765} (2017), 339-343
doi:10.1016/j.physletb.2016.12.029
[arXiv:1512.07538 [astro-ph.CO]].
%11 citations counted in INSPIRE as of 01 Sep 2022

%\cite{Wang:2018caj}
\bibitem{Wang:2018caj}
Y.~T.~Wang, Y.~Cai and Y.~S.~Piao,
``Phase-transition sound of inflation at gravitational waves detectors,''
Phys. Lett. B \textbf{789} (2019), 191-196
%doi:10.1016/j.physletb.2018.12.032
[arXiv:1801.03639 [astro-ph.CO]].
%8 citations counted in INSPIRE as of 01 Sep 2022

%\cite{An:2020fff}
\bibitem{An:2020fff}
H.~An, K.~F.~Lyu, L.~T.~Wang and S.~Zhou,
``A unique gravitational wave signal from phase transition during inflation,''
[arXiv:2009.12381 [astro-ph.CO]].
%5 citations counted in INSPIRE as of 09 Feb 2022

%\cite{Li:2020cjj}
\bibitem{Li:2020cjj}
H.~H.~Li, G.~Ye and Y.~S.~Piao,
``Is the NANOGrav signal a hint of dS decay during inflation?,''
Phys. Lett. B \textbf{816} (2021), 136211
%doi:10.1016/j.physletb.2021.136211
[arXiv:2009.14663 [astro-ph.CO]].
%25 citations counted in INSPIRE as of 01 Sep 2022

%\cite{An:2022cce}
\bibitem{An:2022cce}
H.~An, K.~F.~Lyu, L.~T.~Wang and S.~Zhou,
``Gravitational Waves from an Inflation Triggered First-Order Phase Transition,''
[arXiv:2201.05171 [astro-ph.CO]].
%0 citations counted in INSPIRE as of 09 Feb 2022

%\cite{Sugimura:2011tk}
\bibitem{Sugimura:2011tk}
K.~Sugimura, D.~Yamauchi and M.~Sasaki,
``Multi-field open inflation model and multi-field dynamics in tunneling,''
JCAP \textbf{01} (2012), 027
%doi:10.1088/1475-7516/2012/01/027
[arXiv:1110.4773 [gr-qc]].
%22 citations counted in INSPIRE as of 07 Sep 2022

%\cite{Planck:2018nkj}
\bibitem{Planck:2018nkj}
N.~Aghanim \textit{et al.} [Planck],
``Planck 2018 results. I. Overview and the cosmological legacy of Planck,''
Astron. Astrophys. \textbf{641} (2020), A1
%doi:10.1051/0004-6361/201833880
[arXiv:1807.06205 [astro-ph.CO]].
%809 citations counted in INSPIRE as of 09 Feb 2022

%\cite{Planck:2018lbu}
\bibitem{Planck:2018lbu}
N.~Aghanim \textit{et al.} [Planck],
``Planck 2018 results. VIII. Gravitational lensing,''
Astron. Astrophys. \textbf{641} (2020), A8
%doi:10.1051/0004-6361/201833886
[arXiv:1807.06210 [astro-ph.CO]].
%428 citations counted in INSPIRE as of 31 Aug 2022

%\cite{Baker:2019ndr}
\bibitem{Baker:2019ndr}
M.~J.~Baker, J.~Kopp and A.~J.~Long,
``Filtered Dark Matter at a first-order Phase Transition,''
Phys. Rev. Lett. \textbf{125} (2020) no.15, 151102
%doi:10.1103/PhysRevLett.125.151102
[arXiv:1912.02830 [hep-ph]].
%41 citations counted in INSPIRE as of 04 Apr 2022

%\cite{Chway:2019kft}
\bibitem{Chway:2019kft}
D.~Chway, T.~H.~Jung and C.~S.~Shin,
``Dark matter filtering-out effect during a first-order phase transition,''
Phys. Rev. D \textbf{101} (2020) no.9, 095019
%doi:10.1103/PhysRevD.101.095019
[arXiv:1912.04238 [hep-ph]].
%30 citations counted in INSPIRE as of 04 Apr 2022

%\cite{Marfatia:2020bcs}
\bibitem{Marfatia:2020bcs}
D.~Marfatia and P.~Y.~Tseng,
``Gravitational wave signals of dark matter freeze-out,''
JHEP \textbf{02} (2021), 022
%doi:10.1007/JHEP02(2021)022
[arXiv:2006.07313 [hep-ph]].
%12 citations counted in INSPIRE as of 04 Apr 2022

%\cite{Baldes:2020kam}
\bibitem{Baldes:2020kam}
I.~Baldes, Y.~Gouttenoire and F.~Sala,
``String Fragmentation in Supercooled Confinement and Implications for Dark Matter,''
JHEP \textbf{04} (2021), 278
%doi:10.1007/JHEP04(2021)278
[arXiv:2007.08440 [hep-ph]].
%20 citations counted in INSPIRE as of 04 Apr 2022

%\cite{Azatov:2021ifm}
\bibitem{Azatov:2021ifm}
A.~Azatov, M.~Vanvlasselaer and W.~Yin,
``Dark Matter production from relativistic bubble walls,''
JHEP \textbf{03} (2021), 288
%doi:10.1007/JHEP03(2021)288
[arXiv:2101.05721 [hep-ph]].
%21 citations counted in INSPIRE as of 04 Apr 2022

%\cite{Bian:2021vmi}
\bibitem{Bian:2021vmi}
L.~Bian, X.~Liu and K.~P.~Xie,
``Probing superheavy dark matter with gravitational waves,''
JHEP \textbf{11} (2021), 175
%doi:10.1007/JHEP11(2021)175
[arXiv:2107.13112 [hep-ph]].
%6 citations counted in INSPIRE as of 04 Apr 2022

%\cite{Freese:2022qrl}
\bibitem{Freese:2022qrl}
K.~Freese and M.~W.~Winkler,
``Have Pulsar Timing Arrays detected the Hot Big Bang? Gravitational Waves from Strong First Order Phase Transitions in the Early Universe,''
[arXiv:2208.03330 [astro-ph.CO]].
%0 citations counted in INSPIRE as of 05 Sep 2022

%\cite{Caldwell:2022qsj}
\bibitem{Caldwell:2022qsj}
R.~Caldwell, Y.~Cui, H.~K.~Guo, V.~Mandic, A.~Mariotti, J.~M.~No, M.~J.~Ramsey-Musolf, M.~Sakellariadou, K.~Sinha and L.~T.~Wang, \textit{et al.}
``Detection of Early-Universe Gravitational Wave Signatures and Fundamental Physics,''
[arXiv:2203.07972 [gr-qc]].
%21 citations counted in INSPIRE as of 07 Sep 2022

%\cite{Traschen:1990sw}
\bibitem{Traschen:1990sw}
J.~H.~Traschen and R.~H.~Brandenberger,
``Particle Production During Out-of-equilibrium Phase Transitions,''
Phys. Rev. D \textbf{42} (1990), 2491-2504
%doi:10.1103/PhysRevD.42.2491
%873 citations counted in INSPIRE as of 27 Aug 2022

%\cite{Kofman:1997yn}
\bibitem{Kofman:1997yn}
L.~Kofman, A.~D.~Linde and A.~A.~Starobinsky,
``Towards the theory of reheating after inflation,''
Phys. Rev. D \textbf{56} (1997), 3258-3295
%doi:10.1103/PhysRevD.56.3258
[arXiv:hep-ph/9704452 [hep-ph]].
%1740 citations counted in INSPIRE as of 27 Aug 2022

%\cite{Parry:1998pn}
\bibitem{Parry:1998pn}
M.~Parry and R.~Easther,
``Preheating and the Einstein field equations,''
Phys. Rev. D \textbf{59} (1999), 061301
%doi:10.1103/PhysRevD.59.061301
[arXiv:hep-ph/9809574 [hep-ph]].
%56 citations counted in INSPIRE as of 27 Aug 2022

%\cite{Sornborger:1998mh}
\bibitem{Sornborger:1998mh}
A.~Sornborger and M.~Parry,
``Patterns from Preheating,''
Phys. Rev. Lett. \textbf{83} (1999), 666-669
%doi:10.1103/PhysRevLett.83.666
[arXiv:hep-ph/9811520 [hep-ph]].
%10 citations counted in INSPIRE as of 27 Aug 2022

%\cite{Punturo:2010zz}
\bibitem{Punturo:2010zz}
M.~Punturo, M.~Abernathy, F.~Acernese, B.~Allen, N.~Andersson, K.~Arun, F.~Barone, B.~Barr, M.~Barsuglia and M.~Beker, \textit{et al.}
``The Einstein Telescope: A third-generation gravitational wave observatory,''
Class. Quant. Grav. \textbf{27} (2010), 194002
%doi:10.1088/0264-9381/27/19/194002
%1060 citations counted in INSPIRE as of 31 Aug 2022

%\cite{Reitze:2019iox}
\bibitem{Reitze:2019iox}
D.~Reitze, R.~X.~Adhikari, S.~Ballmer, B.~Barish, L.~Barsotti, G.~Billingsley, D.~A.~Brown, Y.~Chen, D.~Coyne and R.~Eisenstein, \textit{et al.}
%``Cosmic Explorer: The U.S. Contribution to Gravitational-Wave Astronomy beyond LIGO,''
Bull. Am. Astron. Soc. \textbf{51} (2019) no.7, 035
[arXiv:1907.04833 [astro-ph.IM]].
%452 citations counted in INSPIRE as of 31 Aug 2022

%\cite{Cai:2019cdl}
\bibitem{Cai:2019cdl}
R.~G.~Cai, S.~Pi and M.~Sasaki,
``Universal infrared scaling of gravitational wave background spectra,''
Phys. Rev. D \textbf{102} (2020) no.8, 083528
%doi:10.1103/PhysRevD.102.083528
[arXiv:1909.13728 [astro-ph.CO]].
%74 citations counted in INSPIRE as of 07 Sep 2022

%\cite{Caprini:2009fx}
\bibitem{Caprini:2009fx}
C.~Caprini, R.~Durrer, T.~Konstandin and G.~Servant,
``General Properties of the Gravitational Wave Spectrum from Phase Transitions,''
Phys. Rev. D \textbf{79} (2009), 083519
%doi:10.1103/PhysRevD.79.083519
[arXiv:0901.1661 [astro-ph.CO]].
%193 citations counted in INSPIRE as of 07 Sep 2022

%\cite{Huber:2008hg}
\bibitem{Huber:2008hg}
S.~J.~Huber and T.~Konstandin,
``Gravitational Wave Production by Collisions: More Bubbles,''
JCAP \textbf{09} (2008), 022
%doi:10.1088/1475-7516/2008/09/022
[arXiv:0806.1828 [hep-ph]].
%349 citations counted in INSPIRE as of 27 Aug 2022

%\cite{Gouttenoire:2021jhk}
\bibitem{Gouttenoire:2021jhk}
Y.~Gouttenoire, G.~Servant and P.~Simakachorn,
``Kination cosmology from scalar fields and gravitational-wave signatures,''
[arXiv:2111.01150 [hep-ph]].
%18 citations counted in INSPIRE as of 20 Aug 2022

%\cite{Spokoiny:1993kt}
\bibitem{Spokoiny:1993kt}
B.~Spokoiny,
``Deflationary universe scenario,''
Phys. Lett. B \textbf{315} (1993), 40-45
%doi:10.1016/0370-2693(93)90155-B
[arXiv:gr-qc/9306008 [gr-qc]].
%263 citations counted in INSPIRE as of 20 Aug 2022

%\cite{Peebles:1998qn}
\bibitem{Peebles:1998qn}
P.~J.~E.~Peebles and A.~Vilenkin,
``Quintessential inflation,''
Phys. Rev. D \textbf{59} (1999), 063505
%doi:10.1103/PhysRevD.59.063505
[arXiv:astro-ph/9810509 [astro-ph]].
%595 citations counted in INSPIRE as of 20 Aug 2022

%\cite{Banks:2010zn}
\bibitem{Banks:2010zn}
T.~Banks and N.~Seiberg,
``Symmetries and Strings in Field Theory and Gravity,''
Phys. Rev. D \textbf{83} (2011), 084019
%doi:10.1103/PhysRevD.83.084019
[arXiv:1011.5120 [hep-th]].
%618 citations counted in INSPIRE as of 01 Aug 2022

%\cite{Harlow:2018tng}
\bibitem{Harlow:2018tng}
D.~Harlow and H.~Ooguri,
``Symmetries in quantum field theory and quantum gravity,''
Commun. Math. Phys. \textbf{383} (2021) no.3, 1669-1804
%doi:10.1007/s00220-021-04040-y
[arXiv:1810.05338 [hep-th]].
%239 citations counted in INSPIRE as of 07 Aug 2022

%\cite{Harlow:2020bee}
\bibitem{Harlow:2020bee}
D.~Harlow and E.~Shaghoulian,
``Global symmetry, Euclidean gravity, and the black hole information problem,''
JHEP \textbf{04} (2021), 175
%doi:10.1007/JHEP04(2021)175
[arXiv:2010.10539 [hep-th]].
%55 citations counted in INSPIRE as of 07 Aug 2022

%\cite{Gilbert:1989nq}
\bibitem{Gilbert:1989nq}
G.~Gilbert,
``WORMHOLE INDUCED PROTON DECAY,''
Nucl. Phys. B \textbf{328}, 159-170 (1989)
%doi:10.1016/0550-3213(89)90097-7
%121 citations counted in INSPIRE as of 28 Jul 2022

%\cite{Krauss:1988zc}
\bibitem{Krauss:1988zc}
L.~M.~Krauss and F.~Wilczek,
``Discrete Gauge Symmetry in Continuum Theories,''
Phys. Rev. Lett. \textbf{62}, 1221 (1989)
%doi:10.1103/PhysRevLett.62.1221
%660 citations counted in INSPIRE as of 28 Jul 2022

%\cite{LISA:2017pwj}
\bibitem{LISA:2017pwj}
P.~Amaro-Seoane \textit{et al.} [LISA],
``Laser Interferometer Space Antenna,''
[arXiv:1702.00786 [astro-ph.IM]].
%1776 citations counted in INSPIRE as of 11 Aug 2022

%\cite{eLISA:2013xep}
\bibitem{eLISA:2013xep}
P.~A.~Seoane \textit{et al.} [eLISA],
``The Gravitational Universe,''
[arXiv:1305.5720 [astro-ph.CO]].
%341 citations counted in INSPIRE as of 11 Aug 2022

%\cite{Kawamura:2011zz}
\bibitem{Kawamura:2011zz}
S.~Kawamura, M.~Ando, N.~Seto, S.~Sato, T.~Nakamura, K.~Tsubono, N.~Kanda, T.~Tanaka, J.~Yokoyama and I.~Funaki, \textit{et al.}
``The Japanese space gravitational wave antenna: DECIGO,''
Class. Quant. Grav. \textbf{28} (2011), 094011
%doi:10.1088/0264-9381/28/9/094011
%397 citations counted in INSPIRE as of 11 Aug 2022

%\cite{Yagi:2011wg}
\bibitem{Yagi:2011wg}
K.~Yagi and N.~Seto,
``Detector configuration of DECIGO/BBO and identification of cosmological neutron-star binaries,''
Phys. Rev. D \textbf{83} (2011), 044011
[erratum: Phys. Rev. D \textbf{95} (2017) no.10, 109901]
%doi:10.1103/PhysRevD.83.044011
[arXiv:1101.3940 [astro-ph.CO]].
%262 citations counted in INSPIRE as of 11 Aug 2022

%\cite{Harry:2006fi}
\bibitem{Harry:2006fi}
G.~M.~Harry, P.~Fritschel, D.~A.~Shaddock, W.~Folkner and E.~S.~Phinney,
``Laser interferometry for the big bang observer,''
Class. Quant. Grav. \textbf{23} (2006), 4887-4894
[erratum: Class. Quant. Grav. \textbf{23} (2006), 7361]
%doi:10.1088/0264-9381/23/15/008
%191 citations counted in INSPIRE as of 07 May 2022

%\cite{Corbin:2005ny}
\bibitem{Corbin:2005ny}
V.~Corbin and N.~J.~Cornish,
``Detecting the cosmic gravitational wave background with the big bang observer,''
Class. Quant. Grav. \textbf{23} (2006), 2435-2446
%doi:10.1088/0264-9381/23/7/014
[arXiv:gr-qc/0512039 [gr-qc]].
%233 citations counted in INSPIRE as of 11 Aug 2022

%\cite{TianQin:2015yph}
\bibitem{TianQin:2015yph}
J.~Luo \textit{et al.} [TianQin],
``TianQin: a space-borne gravitational wave detector,''
Class. Quant. Grav. \textbf{33} (2016) no.3, 035010
%doi:10.1088/0264-9381/33/3/035010
[arXiv:1512.02076 [astro-ph.IM]].
%585 citations counted in INSPIRE as of 11 Aug 2022 

%\cite{Ruan:2018tsw}
\bibitem{Ruan:2018tsw}
W.~H.~Ruan, Z.~K.~Guo, R.~G.~Cai and Y.~Z.~Zhang,
``Taiji program: Gravitational-wave sources,''
Int. J. Mod. Phys. A \textbf{35} (2020) no.17, 2050075
%doi:10.1142/S0217751X2050075X
[arXiv:1807.09495 [gr-qc]].
%219 citations counted in INSPIRE as of 11 Aug 2022

%\cite{Yang:2022quy}
\bibitem{Yang:2022quy}
J.~Yang, R.~Zhou and L.~Bian,
``Gravitational waves and monopoles dark matter from first-order phase transition,''
[arXiv:2204.07540 [hep-ph]].
%0 citations counted in INSPIRE as of 27 Aug 2022


\end{thebibliography}
\end{document}